\documentclass[nopacs,nokeys]{revtex4}%
\usepackage{amsfonts}
\usepackage{amsmath}
\usepackage{amssymb}
\usepackage{bbm}
\usepackage{bm}
\usepackage[utf8]{inputenc}
\usepackage{tensor}
\usepackage{slashed}
\usepackage[centertableaux ]{ytableau}
\usepackage{hyperref}
\usepackage{natbib}
\usepackage{enumerate}
\usepackage{braket,mleftright}
\usepackage{graphicx}

\usepackage{tcolorbox}
\usepackage{color}
 \definecolor{jblue}  {RGB}{20,50,100}
  \definecolor{npurple}  {RGB} {153, 51, 204}
  \definecolor{wred}   {RGB}{217,0,56}
  \definecolor{white}   {RGB}{255,255,255}
  \definecolor{korange}   {RGB}{235, 80,  43}
  \definecolor{korange2}   {RGB}{245, 100,  63}
  \definecolor{kyelloworange}   {RGB}{255, 210,  110}
  \definecolor{kyelloworange2}   {RGB}{240, 170,  90}
  \definecolor{kred}   {RGB}{204,  102, 153}
  \definecolor{kpurple}   {RGB}{153,  61, 190}
  \definecolor{kpurplelight}   {RGB}{213,  161, 230}
 \definecolor{nblue}   {RGB}{200,  230, 250}
\definecolor{mycolor}{RGB}{5,61,245}

\begin{document}

\title{Complete analytical solution to the Cornell potential and heavy quarkonium structure}
 \author{ M. Napsuciale$^{(1)}$, S. Rodr\'{\i}guez$^{(2)}$,  A.E. Villanueva-Guti\'{e}rrez$^{(3)}$}
\address{$^{(1)}$Departamento de F\'{i}sica, Universidad de Guanajuato, Lomas del Campestre 103, Fraccionamiento
Lomas del Campestre, 37150, Le\'on, Guanajuato, M\'{e}xico.}
\address{$^{(2)}$Facultad de Ciencias F\'isico-Matem\'{a}ticas,
  Universidad Aut\'onoma de Coahuila, Edificio A, Unidad
  Camporredondo, 25000, Saltillo, Coahuila, M\'exico.} 
\address{$^{(3)}$Escuela Superior de F\'{i}sica y Matem\'{a}ticas, Instituto Polit\'{e}cnico Nacional,
Avenida Instituto Polit\'{e}cnico Nacional S/N, Edificio 9, Unidad Profesional “Adolfo L\'{o}pez Mateos”, Colonia San Pedro Zacatenco,
 Delegaci\'{o}n Gustavo A. Madero, 07700, CDMX, México} 

\begin{abstract}
We use the recently proposed supersymmetric expansion algorithm (SEA) to obtain a complete analytical solution to the Schr\"{o}dinger 
equation with the Cornell potential. We find that the energy levels $E_{nl}(\lambda)$ depend on $n^{2}$ and $L^{2}=l(l+1)$.  
For a given $n$, the energy {\emph {decreases}} with $l$ and the radial probabilities have the Coulomb shape but their peaks are 
shifted toward smaller radius. We study the heavy quarkonium structure on the light of these results, showing that the measured 
$\bar{b}b$ and $\bar{c}c$ meson masses follow the inverted spectrum pattern predicted by the Cornell potential. Details of the 
structure of heavy quarkonium like the mean inverse radius and mean squared velocity for the different quarkonium configurations 
can be obtained from our solution. These details point to significant relativistic corrections for all the configurations of real heavy quarkonium.
We calculate relativistic corrections using perturbation theory finding an expansion in $\alpha^{2}_{s}$ for the heavy quarkonium masses.
The mass hierarchies in the fine splittings can be qualitatively understood 
from this expansion. The quantitative analysis of the Bohr-like levels and of the fine splittings in the $l=0$ sector allow us to make well 
defined predictions for the masses of some of the missing heavy quarkonium states, to identify the $\psi(4040)$ as the $3^{3}S_{1}$ $\bar{c}c$ 
state and the $\psi(3842)$, $\psi(3823)$ and  $\psi(3770)$ as the $3^{3}D_{3}$, $3^{3}D_{2}$ and $3^{3}D_{1}$ $\bar{c}c$ 
states respectively.
\end{abstract}

\maketitle

\section{Introduction}

The Cornell potential, also named linear plus Coulomb or funnel potential in the literature, was one of the first models for the 
phenomenological description of the confined dynamics of quarkonium \cite{Eichten:1974af, Eichten:1978tg, Eichten:1979ms}. 
The potential is a sum of a Coulomb term and a linear term.  The Coulomb 
term is motivated form the perturbative short distance regime of Quantum Chromodynamics (QCD). At intermediate and long 
distances (compared to the typical size of the heavy quarkonium), we face the unsolved problem of confinement. 

The formal derivation of quark-antiquark forces from QCD started with the pioneering work by Wilson \cite{Wilson:1974sk} who related 
the interquark potential to the so called Wilson loop. The evaluation of the Wilson loop in the long distance 
regime, with diverse methods, yields a linear potential as the leading term 
\cite{Eichten:1979pu, Eichten:1980mw, Buchmuller:1981fr, Gromes:1983pm, Gromes:1984ma, Barchielli:1986zs, Barchielli:1988zp,Brambilla:1997kz}. 
The physical picture that arises at long distances under reasonable 
assumptions is an inter-quark potential generated by a chromo-electric flux tube \cite{Gromes:1983pm}  which produces a QCD string. 
These results put on a firm basis the Coulomb plus linear potential as a suitable starting point for the phenomenological description 
of the structure of heavy quarkonium. 

The systematic calculation of heavy quarkonium properties from QCD had considerable advances with the formulation of effective field 
theories for QCD in the non-relativistic regime. The non-relativistic expansion of QCD ($NRQCD$) \cite{Bodwin:1994jh} and the potential 
non-relativistic QCD expansion ($pNRQCD$) \cite{Brambilla:1999xf}, take care of the scales of heavy quarkonium, integrate physics 
at high energy and yield systematic expansions
in terms of operators suppressed according to the corresponding power counting rules. In the case of $NRQCD$ there are two scales, 
$m_{Q}$ and $m_{Q}v^{2}$, with a well defined hierarchy $m_{Q}>>m_{Q}v^{2}$, and the effective theory is obtained integrating out 
the degrees of freedom at the scale $m_{Q}$. For heavy quarkonium, a hierarchy $m_{Q}>>m_{Q}v>>m_{Q}v^{2}>>\Lambda_{QCD}$  
is expected and $pNRQCD$ obtains the effective theory for physics at the ultrasoft scale $m_{Q}v^{2}$ integrating out also physics at the 
soft scale $m_{Q}v$. A primary concern for the $pNRQCD$ is the validity of this scale hierarchy which grants the validity of the 
perturbative matching which otherwise must be done considering the strong coupling regime. Presently, it is not clear if all heavy quarkonium 
configurations respect the above hierarchy (the weak coupling scenario) or for some states actually it is necessary to work in the strong 
coupling matching regime.
 On the other side, calculations in the lattice became efficient with the use of effective field theories and yield the same conclusion: 
the non-relativistic description of QCD in the non-perturbative region is given by the Cornell potential \cite{Koma:2006si,Koma:2007jq}.
A complete analytical solution to the Cornell potential is desirable because it can shed light into the structure of real 
heavy quarkonium, specially of the relative size of the soft and ultrasoft scales and $\Lambda_{QCD}$.

The approach followed in Refs. \cite{Eichten:1974af, Eichten:1978tg, Eichten:1979ms} and almost in every subsequent work on 
analytical solutions to the Cornell potential, was to consider first the linear potential whose exact solutions for $S$-waves are given in terms  
of the Airy functions, and to incorporate later the effects of the Coulomb part using Rayleigh-Schrodinger perturbation theory to obtain 
approximate solutions with an expansion in powers of the strong coupling constant $\alpha_{s}$. The lack of complete analytical 
solutions to the Cornell potential motivated the use of general results aiming to guess from non-relativistic quantum mechanics the structure of 
heavy quarkonium \cite{Quigg:1979vr,Lucha:1991vn}, but most of the work on the field has been done using numerical solutions for the 
Cornell potential \cite{Godfrey:1985xj,Lucha:1998xc,Barnes:2005pb,Domenech-Garret:2008zbm,Li:2009zu,Leitao:2014jha,Godfrey:2015dia,Godfrey:2015dva,Deng:2016stx,Deng:2016ktl, Soni:2017wvy, Mateu:2018zym}. 

Although such solutions have been used for almost half a century to get 
insight on the heavy quarkonium dynamics, the relative importance of the non-perturbative effects modeled by the linear 
term and the perturbative effects has not been clearly stablished, precisely because the later are incorporated only in a perturbative 
manner. Recently, there has been renewed interest 
in the heavy quarkonium spectrum due to the possibilities opened by new factories for the discovery of conventional heavy quarkonium 
as well as exotic states (for a review on these possibilities and a more complete list of references to old literature see  \cite{Eichten:2007qx,Brambilla:2010cs,Patrignani:2012an}).

A review and classification of the different methods aiming to obtain analytical solutions to the Schr\"{o}dinger equation and a novel 
systematic method named Supersymmetric Expansion Algorithm (SEA) was recently published \cite{Napsuciale:2024yrf}. 
The SEA allows us to use the full power of supersymmetry in non-relativistic quantum mechanics and complete 
analytical solutions to long standing unsolved potentials relevant to several branches of natural sciences like the Yukawa potential \cite{Napsuciale:2020ehf}, \cite{Napsuciale:2021qtw}, Hulth\'{e}n and anharmonic potentials \cite{Napsuciale:2024yrf} have been obtained by this method.
In this work, we use the SEA to obtain the complete analytical solutions of the Cornell potential, 
apply the solutions to the description of the physics of heavy quarkonium and explore the leading relativistic corrections.
We are able to extract general qualitative and quantitative predictions for the bottomonium and charmonium spectrum which nicely 
agree with existing data.  

This paper is organized as follows. In the next section we give the detail of the calculation of the complete analytical solution for the Cornell
potential. Section III is devoted to the calculation of the heavy quarkonium spectrum using the solutions to the Cornell potential and its 
comparison with existing data.  In section IV we consider the leading relativistic corrections that yield the fine splittings. Section V is devoted 
to a first quantitative analysis of the fine splittings in the $l=0$ sector, qualitative analysis of the fine splittings for $l=1,2$ and the corresponding 
quantitative and qualitative predictions for the bottomonium and charmonium spectrum. We close with our conclusions and perspectives 
in section VI.       

\section{The Cornell potential in the SEA}

The Cornell potential is given by
\begin{equation}
V_{c}(r)= -\frac{\alpha \hbar c}{r} +  \frac{\sigma}{\hbar c} r.
\end{equation}
We introduce in this section the $\hbar$ and $c$ factor in order to have a dimensionless parameter $\alpha$ and a parameter 
$\sigma$ with units of $E^{2}$. In the remaining of this section we will solve the Schr\"odinger equation (SE) for the linear plus Coulomb 
potential using the supersymmetric expansion algorithm introduced in Ref. \cite{Napsuciale:2024yrf}.
We refer the reader to that work for further details of the formalism.  In a first step, we use the typical distance scale 
of the system as the scale related to the coupling $\alpha$, {\it{i.e.}} the Bohr radius 
\begin{equation}
a=\hbar/\mu c\alpha, 
\label{a0}
\end{equation}
where $\mu$ is the reduced mass of the system, to cast the radial part of the Schr\"{o}dinger equation into its dimensionless form 
\begin{equation}
\left[-\frac{d^{2}}{dx^{2}}+v_{0}(x,\lambda)\right] u_{0}(x,\lambda)=\epsilon_{0}(\lambda)u_{0}(x,\lambda),
\end{equation}
where a suffix $``0"$ is attached to all quantities because this is the initial step of the algorithm. Here, $x=r/a$, $R(r)=u_{0}(x)/x$, 
$\epsilon_{0}(\lambda)=E_{0}(\lambda)/E_{c}$, with the typical energy scale given by the Coulomb energy $E_{c}=\hbar^{2}/2\mu a^{2}=\mu c^{2}\alpha^{2}/2$. 
The dimensionless effective potential for the Cornell potential is given by
\begin{equation}
v_{0}(x,\lambda)= \frac{l(l+1)}{x^{2}} -\frac{2}{x}+\lambda x,
\end{equation}
with the normalized string tension
\begin{equation}
\lambda=\frac{\sigma}{E_{c}E_{s}}=\frac{2\sigma}{(\mu c^{2})^{2}\alpha^{3}},
\label{lambda}
\end{equation}
where $E_{s}=\hbar c/a=\mu c^{2}\alpha$.

The SEA construct the complete analytical solution starting with very special states named edge states, which in general are 
nodeless (excited or ground) states. With this aim, we first recast the SE into a logarithmic form
\begin{equation}
W^{2}_{0}(x,\lambda)-W_{0}^{\prime}=v_{0}(x,\lambda)-\epsilon_{0}(\lambda).
\label{R0}
\end{equation}
where
\begin{equation}
W_{0}(x,\lambda)=-\frac{d}{dx} \ln u_{0}(x,\lambda).
\end{equation}
The solution to equation (\ref{R0}) is obtained as power series in $\lambda$ 
\begin{equation}
v_{0}(x, \lambda)= \sum_{k=0}^{\infty} v_{0k}(x)\lambda^{k},\qquad W_{0}(x,\lambda)=\sum_{k=0}^{\infty} w_{0k}(x)\lambda^{k},\qquad  
\epsilon_{0}(\lambda)=\sum_{k=0}^{\infty} \varepsilon_{0k}\lambda^{k}.
\end{equation}
The coefficients for the expansion of the Cornell potential in powers of $\lambda$ are given by 
\begin{align}
v_{0k}(x) =\left \{
\begin{array}{r}
\frac{l(l+1)}{x^{2}}-\frac{2}{x} \quad \textit{for} \quad k=0,\\ 
x  \quad \textit{for} \quad k= 1 , \\
0 \quad \textit{for} \quad k\ge 2.
  \end{array}
  \right.
\end{align}
The coefficients in the power series of $W_{0}(x,\lambda)$ and $\epsilon_{0}(\lambda)$ satisfy the following infinite set of hierarchical 
equations 
\begin{align}
k&=0:  & w_{00}^{2} - w^{\prime}_{00} &= \frac{l(l+1)}{x^{2}}-\frac{2}{x} - \varepsilon_{00},  \\
k&=1:  &2w_{00} w_{01} - w^{\prime}_{01} &= x -\varepsilon_{01} ,\\
k&\ge 2:  & 2w_{00} w_{0k} - w^{\prime}_{0k} &= - B_{0k}(x) -\varepsilon_{0k},  \label{w0k}
\end{align}
where
\begin{equation}
B_{0k}=\sum_{m+n=k}w_{0m}w_{0n}.
\end{equation}
The solutions up to $k=5$ are given by
\begin{align}
w_{00}&=\frac{1}{b}-\frac{b}{x}, \\
w_{01}&= \frac{b}{2} x, \\
w_{02}&= -\frac{b^{3}}{8} \left[ b (b+1) x + x^{2} \right],\\
w_{03}&= \frac{b^{5}}{32} \left[ b^{2} (b+1)(4b +5)x +b(4b+5)x^{2} +2 x^{3} \right],  \\
w_{04}&=-\frac{b^{7}}{256} \left[ b^{3}(b+1) (48 b^{2} +129 b + 88)x + b^{2}(48 b^{2} +129 b + 88) x^{2} + 2b(15b+22) x^{3} +10 x^{4} \right] , \\
w_{05}&= \frac{b^{9}}{512} \left[ b^{4}(b+1) (176 b^{3} + 753 b^{2} + 1049 b + 539)x + b^{3} (176 b^{3} + 753 b^{2} + 1049 b + 539)x^{2}  \right. \nonumber \\
&\left. + 2 b^{2}(60 b^{2} + 185 b + 147) x^{3}  + b (56 b + 93)x^{4} + 14 x^{5} \right] , \\
\varepsilon_{00}&=-\frac{1}{b^{2}}, \\
\varepsilon_{01}&=\frac{b}{2}(2b +1), \\
\varepsilon_{02}&= -\frac{b^{4}}{8}(b+1)(2b+1), \\
\varepsilon_{03}&= \frac{b^{7}}{32} (b+1)(2b +1) (4b +5), \\
\varepsilon_{04}&=-\frac{b^{10}}{256} (b+1)(2b +1) (48b^{2}+129b+88), \\
\varepsilon_{05}&=\frac{b^{13}}{512} (b+1)(2b +1) (176 b^{3}+753 b^{2} + 1049 b + 539), 
\end{align}
where
\begin{equation}
b=l+1.
\end{equation}
In general, for $k\ge 2$ we have polynomial solutions for $w_{0k}(x)$
\begin{equation}
w_{0k}(x)=\sum_{\alpha=1}^{k} w_{0k\alpha}x^{\alpha}
\end{equation} 
and inserting this expression in Eqs. (\ref{w0k}) we get the following recursion relations for the constant coefficients $w_{0k\alpha}$
\begin{align}
w_{0kk}&=-\frac{b}{2}B_{0kk}, \\
w_{0k\alpha}&=\frac{b}{2}\left(  \left(  2b+\alpha+1\right) w_{0k(\alpha+1)}-B_{0k\alpha}\right), \qquad \alpha=k-1,k-2,...3,2, \\
\varepsilon_{0k}&=\left(  2b+1\right)  w_{0k1},
\end{align}
where
\begin{equation}
B_{0k\alpha}=\sum_{m+n=k}\, \sum_{\beta+\gamma=\alpha} w_{0m\beta}w_{0n\gamma}.
\end{equation}
These recurrence relations yields the first solution for the Cornell Hamiltonian 
\begin{equation}
H_{0}=-\frac{d^{2}}{dx^{2}}+v_{0}(x,\lambda).
\end{equation}
Indeed, separating the $\lambda$-independent terms
\begin{equation}
W_{0}(x,\lambda)=\frac{1}{l+1}-\frac{l+1}{x} + \sum_{k=1}^{\infty}w_{0k}(x)\lambda^{k},
\end{equation}
we can see that the $\lambda$-independent terms correspond to the solution to the Coulomb problem, thus, for this solution, $l+1=n$ and 
skipping normalization factors we get
\begin{align}
u_{0}(x,\lambda)&= x^{n}e^{-\frac{x}{n}} e^{-G_{0}(x,\lambda)}\equiv \phi^{(0)}_{n,n-1}(x,\lambda), \\
\epsilon_{0}(\lambda)&=-\frac{1}{n^{2}}+\sum_{k=1}^{\infty} \left( 2n+1\right)w_{0k1} \lambda^{k}, \label{en0cor}
\end{align}
where we used $\phi^{(0)}_{nl}(x,\lambda)$ for the solutions with the principal quantum number $n$ and angular momentum $l$ of the Cornell 
Hamiltonian $H_{0}$. The function $G_{0}$ is given by
\begin{equation}
G_{0}(x,\lambda)= \sum_{k=1}^{\infty}\lambda^{k}\left( \sum_{\alpha=1}^{k}w_{0k\alpha} \frac{x^{\alpha+1}}{\alpha+1} \right).
\end{equation}
The solution $\phi^{(0)}_{n, n-1}(x,\lambda)$ is the edge state for the $n$-th level of the Cornel Hamiltonian $H_{0}$. The rest of the states in this level 
are constructed with the aid of supersymmetry. We start noticing that $H_{0}$ can be factorized as
\begin{equation}
H_{0}= a^{\dagger}_{0}a_{0}+\epsilon_{0},
\end{equation}
where
\begin{equation}
a_{0}=\frac{d}{dx}+W_{0}, \qquad a^{\dagger}_{0}=-\frac{d}{dx}+W_{0},
\end{equation}
and the edge state $u_{0}$ satisfies
\begin{equation}
a_{0} u_{0}(x,\lambda)= 0 .
\end{equation}
Then we construct the supersymmetric partner
\begin{equation}
H_{1}= a_{0} a^{\dagger}_{0}+\epsilon_{0}=-\frac{d^{2}}{dx^{2}}+v_{1}(x,\lambda),
\end{equation}
with
\begin{equation}
v_{1}(x,\lambda)=v_{0}(x,\lambda)+2W^{\prime}_{0}= \frac{l(l+1)}{x^{2}} -\frac{2}{x}+ \lambda x + 2\sum_{k=0}^{\infty}w^{\prime}_{0k}(x)\lambda^{k} .
\end{equation}
Now we find the edge state solution for $H_{1}$ in a similar way as we did it for $H_{0}$, i.e., we recast the Schr\"{o}dinger equation 
for $H_{1}$ in the logarithmic form
\begin{equation}
W^{2}_{1}-W^{\prime}_{1}=v_{1}(x,\lambda)-\epsilon_{1}(\lambda)=v_{0}(x,\lambda)+2W^{\prime}_{0}-\epsilon_{1}(\lambda),
\label{R1}
\end{equation}
where
\begin{equation}
W_{1}(x,\lambda)=-\frac{d}{dx} \ln u_{1}(x,\lambda).
\end{equation}
The solution to equation (\ref{R1}) is obtained as power series in $\lambda$ 
\begin{equation}
v_{1}(x, \lambda)= \sum_{k=0}^{\infty} v_{1k}(x)\lambda^{k},\qquad W_{1}(x,\lambda)=\sum_{k=1}^{\infty} w_{1k}(x)\lambda^{k},\qquad  
\epsilon_{1}(\lambda)=\sum_{k=1}^{\infty} \varepsilon_{1k}\lambda^{k}.
\end{equation}
The coefficients for the expansion of $v_{1}$ in powers of $\lambda$ are given by 
\begin{align}
v_{1k}(x) =\left \{
\begin{array}{r}
\frac{l(l+1)}{x^{2}}-\frac{2}{x} +2w^{\prime}_{00}\quad \textit{for} \quad k=0,\\ 
x +2w^{\prime}_{01} \quad \textit{for} \quad k= 1 , \\
2w^{\prime}_{0k} \quad \textit{for} \quad k\ge 2.
  \end{array}
  \right.
\end{align}
The coefficients in the power series of $W_{1}(x,\lambda)$ and $\epsilon_{1}(\lambda)$ satisfy now the following infinite set of hierarchical 
equations 
\begin{align}
k&=0:  & w_{10}^{2} - w^{\prime}_{10} &= \frac{(l+1)(l+2)}{x^{2}}-\frac{2}{x} - \varepsilon_{10},  \\
k&=1:  &2w_{10} w_{11} - w^{\prime}_{11} &= x +2w^{\prime}_{01} -\varepsilon_{11}, \\
k&\ge 2:  & 2w_{10} w_{1k} - w^{\prime}_{1k} &= 2w^{\prime}_{0k}- B_{1k} -\varepsilon_{1k},  \label{w1k}
\end{align}
where
\begin{equation}
B_{1k}=\sum_{m+n=k}w_{1m}w_{1n}.
\end{equation}
The solutions for $k=0,1$ are
\begin{align}
w_{10}&=\frac{1}{b+1}-\frac{b+1}{x}, &\varepsilon_{10}&=-\frac{1}{(b+1)^{2}}, \\
w_{11}&= \frac{b+1}{2} x, & \varepsilon_{11}&=\frac{1}{2}(3(b+1)^{2}-b(b-1)), 
\end{align}
while for $k\ge 2$,  $w_{1k}$ has the general form
\begin{equation}
w_{1k}(x)=\sum_{\alpha=1}^{k}w_{1k\alpha}x^{\alpha},
\end{equation}
where the numerical coefficients satisfy the following recurrence relations
\begin{align}
w_{1kk}&=-\frac{b+1}{2}B_{1kk}, \\
w_{1k\alpha}&=\frac{b+1}{2}\left(  \left(  2b+3+\alpha\right)w_{0k(\alpha+1)}-B_{1k\alpha}\right), \qquad \alpha=k-1,k-2,...3,2, \\
\varepsilon_{1k}&=\left(  2b+3\right)  w_{1k1}+ 2w_{0k1}.
\end{align}
From the $\lambda$-independent part of the solution we identify this solution as a $l=n-2$ solution of the level $n$-th of $H_{1}$. 
Explicitly, the unnormalized solution is given by
\begin{align}
u_{1}(x,\lambda)&= x^{n}e^{-\frac{x}{n}} e^{-G_{1}(x,\lambda)}\equiv \phi^{(1)}_{n,n-2}(x,\lambda), \\
\epsilon_{1}(\lambda)&=-\frac{1}{n^{2}}+\sum_{k=1}^{\infty} \left(  (2b+3) w_{1k1}+ 2w_{0k1} \right) \lambda^{k}. \label{en1cor}
\end{align}
The function $G_{1}$ is given by
\begin{equation}
G_{1}(x,\lambda)= \sum_{k=1}^{\infty}\lambda^{k}\left( \sum_{\alpha=1}^{k}w_{1k\alpha} \frac{x^{\alpha+1}}{\alpha+1} \right).
\end{equation}

Supersymmetry allows us to obtain a second solution to the Cornel Hamiltonian $H_{0}$, from the edge eigenstate $u_{1}$ of $H_{1}$. 
Indeed, $u_{1}$ satisfies
\begin{equation}
H_{1}u_{1}(x,\lambda)=[a_{0}a_{0}^{\dagger}+ \epsilon_{0}] u_{1}(x,\lambda)=\epsilon_{1}(\lambda)u_{1}(x,\lambda).
\end{equation}
Acting with $a_{0}^{\dagger}$ on this equation we get
\begin{equation}
[a_{0}^{\dagger}a_{0}+ \epsilon_{0}] a_{0}^{\dagger} u_{1}(x,\lambda)=H_{0}a_{0}^{\dagger} u_{1}(x,\lambda)
=\epsilon_{1}(\lambda)a_{0}^{\dagger}(\lambda)u_{1}(x,\lambda),
\end{equation}
thus, $a_{0}^{\dagger} u_{1}(x,\lambda)$ is also an eigenstate of $H_{0}$ with eigenvalue $\epsilon_{1}(\lambda)$. This is also a solution 
for the $n$-th level but now with $l=n-2$. The unnormalized solution for $H_{0}$ is
\begin{equation}
\phi^{(0)}_{n,n-2}(x,\lambda)= a^{\dagger}_{0}u_{1}(x,\lambda)=a^{\dagger}_{0}\phi^{(1)}_{n,n-2}(x,\lambda).
\end{equation}
Notice that in the usual $nl$ notation, $\epsilon_{1}(\lambda)=\epsilon_{n, n-2}(\lambda)$, $\epsilon_{0}(\lambda)=\epsilon_{n, n-1}(\lambda)$ and 
$\epsilon_{n, n-1}(\lambda)\neq\epsilon_{n, n-2}(\lambda)$ thus de $l$-degeneracy of the $n$-th level of the Coulomb 
potential is broken by the linear term in the Cornell Hamiltonian $H_{0}$.

We continue this process and construct a new solution to the $n$-th level of $H_{0}$ factorizing now $H_{1}$ as 
\begin{equation}
H_{1}=a^{\dagger}_{1}a_{1} + \epsilon_{1}(\lambda),
\end{equation}
with
\begin{equation}
a_{1}=-\frac{d}{dx} + W_{1}(x,\lambda), \qquad a^{\dagger}_{1}=\frac{d}{dx} + W_{1}(x,\lambda),
\end{equation}
and constructing a superpartner to $H_{1}$ defined as
\begin{equation}
H_{2}=a_{1}a^{\dagger}_{1} + \epsilon_{1}(\lambda)=-\frac{d^{2}}{dx^{2}} + v_{2}(x,\lambda),
\end{equation} 
where
\begin{equation}
v_{2}(x,\lambda)=v_{1}(x,\lambda)+ 2 W^{\prime}_{1}.
\end{equation}
We solve likewise this potential obtaining the unnormalized solution of $H_{0}$ as
\begin{equation}
\phi^{(0)}_{n,n-3}(x,\lambda)= a^{\dagger}_{1} a^{\dagger}_{0}u_{2}(x,\lambda)=a^{\dagger}_{1} a^{\dagger}_{0}\phi^{(2)}_{n,n-3}(x,\lambda),
\end{equation}
where $u_{2}(x,\lambda)$ is the solution to $H_{2}$ with eigenvalue $\epsilon_{2}(\lambda)$ which is a common eigenvalue of $H_{2}, H_{1}, H_{0}$.

Repeating the algorithm, in the step $r$ we solve the $r$-th superpartner $H_{r}$ in the same manner. The solutions for $k=0,1$ in this step are
\begin{align}
w_{r0}&=\frac{1}{b+r}-\frac{b+r}{x}, &\varepsilon_{r0}&=-\frac{1}{(b+r)^{2}}, \\
w_{r1}&= \frac{b+r}{2} x, & \varepsilon_{r1}&=\frac{1}{2}(3(b+r)^{2}-b(b-1)), 
\end{align}
while for $k\ge 2$ the coefficients in the expansion of the $r$-th superpotential $W_{r}$ has the general form
\begin{equation}
w_{rk}(x)=\sum_{\alpha=1}^{k}w_{rk\alpha}x^{\alpha},
\label{wrka}
\end{equation}
where the numerical coefficients satisfy the following recurrence relations
\begin{align}
w_{rkk}&=-\frac{b+r}{2}B_{rkk}, \label{wrkarr1} \\ 
w_{rk\alpha}&=\frac{b+r}{2}\left(  \left(  2b+2r+\alpha+1\right)
w_{0k(\alpha+1)}+2\left(  \alpha+1\right)  \sum_{q=0}^{r-1} w_{qk(\alpha+1)} -B_{rk\alpha}\right), \qquad \alpha=k-1,k-2,...3,2, \label{wrkarr2}\\ 
\varepsilon_{rk}&=\left(  2b+2r+1\right)  w_{rk1}+ 2\sum_{q=0}^{r-1}w_{qk1},
\label{wrkarr3}
\end{align}
with
\begin{equation}
B_{rk\alpha}=\sum_{m+n=k} \quad \sum_{\beta+\gamma=\alpha} w_{rm\beta}w_{rn\gamma}.
\end{equation}
The unnormalized solution for the edge state of $H_{r}$,  is given by
\begin{align}
u_{r}(x,\lambda)&= x^{n}e^{-\frac{x}{n}} e^{-G_{r}(x,\lambda)}, \\
\epsilon_{r}(\lambda)&=-\frac{1}{n^{2}} + \sum_{k=0}^{\infty}\left( \left( 2n+1\right)w_{rk1} + 2\sum_{q=0}^{r-1}w_{qk1} \right)\lambda^{k}  , \label{enrcor}
\end{align}
where
\begin{equation}
G_{r}(x,\lambda)=\int w_{r}(x,\lambda)dx = \sum_{k=1}^{\infty}\lambda^{k}\left( \sum_{\alpha=1}^{k}w_{rk\alpha} \frac{x^{\alpha+1}}{\alpha+1} \right).
\end{equation}
The solution for the Cornell Hamiltonian $H_{0}$ in the $nl$ notation is given by
\begin{align}
\phi^{(0)}_{n,n-1-r}&=a^{\dagger}_{0}a^{\dagger}_{1}...a^{\dagger}_{r-2}a^{\dagger}_{r-1} u_{r}, \\
\epsilon_{n,n-1-r}(\lambda)&=-\frac{1}{n^{2}}+\sum_{k=0}^{\infty} \left( \left( 2n+1\right)w_{rk1} + 2\sum_{q=0}^{r-1}w_{qk1} \right) \lambda^{k}.
\end{align} 
The process terminates for $r=n-1$ when we reach the lowest value $l=0$ and the $n$-th level of the normalized Cornell Hamiltonian $H_{0}$ 
is completely solved. Notice that we started with an arbitrary value of $n$ thus, all the levels of the Cornell potential can be solved with the 
supersymmetric expansion algorithm. 
All the information of the analytic solution resides in the coefficients $w_{rk\alpha}$ of the expansion in Eq. (\ref{wrka}) which satisfy the 
algebraic recurrence relations in 
Eq. (\ref{wrkarr1},\ref{wrkarr2},\ref{wrkarr3}). We wrote a symbolic code to solve these relations to the desired order in $\lambda$.

We find that the energy levels $\epsilon_{nl}(\lambda)$ can be written as 
\begin{equation}
\epsilon_{nl}(\lambda)=\sum_{k=0}^{\infty}\varepsilon_{k}(n^{2},L^{2})\lambda^{k},
\label{epsnl}
\end{equation}
where the coefficients $\varepsilon_{k}$ depend on $n^{2}$ and $L^{2}=l(l+1)$. The coefficients for large $k$ have long expressions thus we 
explicitly write them only up to $k=10$ for future reference
\begin{align}
\varepsilon_{0}(n^{2},L^{2})&= -\frac{1}{n^{2}}, \\
\varepsilon_{1}(n^{2},L^{2})&=\frac{1}{2}(3n^{2}-L^{2}),\\
\varepsilon_{2}(n^{2},L^{2})&=-\frac{n^{2}}{16}(7n^{4}-3L^{4}+5 n^{2}),\\
\varepsilon_{3}(n^{2},L^{2})&=\frac{n^{4}}{64}(33 n^{6} - 7n^{2}L^{4} -10 L^{6} + 75 n^{4}),\\
\varepsilon_{4}(n^{2},L^{2})&=-\frac{n^{6}}{512}(465n^{8} - 99n^{4}L^{4} - 90n^{2}L^{6} - 84L^{8} +2275n^{6} -180 n^{2}L^{4} + 440 n^{4}),\\
\varepsilon_{5} (n^{2},L^{2}) & =\frac{n^{8}}{1024}\left(  1995n^{10}+17340n^{8}%
-3n^{6}\left(  155L^{4}-3803\right)  -91n^{4}L^{4}\left(  4L^{2}+23\right) \right. \nonumber \\
& \left.  -88n^{2}L^{6}\left(  3L^{2}+10\right)  -198L^{10}\right), \\
\varepsilon_{6} (n^{2},L^{2}) & =-\frac{n^{10}}{16385}\left(  77027n^{12}+1060290n^{10}%
-133n^{8}\left(  150L^{4}-11663\right)  \right.  \nonumber \\
& -34n^{6}\left(  426L^{6}+4887L^{4}-5000\right)  -15n^{4}L^{4}\left(
607L^{4}+5612L^{2}+4800\right)  \nonumber\\
& \left.  -364n^{2}L^{8}\left(  17L^{2}+75\right)  -4004L^{12}\right), \\
\varepsilon_{7} (n^{2},L^{2})  & =\frac{n^{12}}{131072}\left(  1608201n^{14}+32473350n^{12}%
-69n^{10}\left(  6698L^{4}-1254497\right)  \right. \nonumber \\
& -66n^{8}\left(  4858L^{6}+93961L^{4}-485630\right)  -19n^{6}L^{4}%
(9525L^{4}+175460L^{2}+402116) \nonumber\\
& -\left.  204n^{4}L^{6}\left(  543L^{4}+6378L^{2}+10000\right)
-24480n^{2}L^{10}\left(  3L^{2}+16\right)  -66432L^{14}\right), \\
\varepsilon_{8}(n^{2},L^{2})   & =-\frac{n^{14}}{2097152}\left(  71016319n^{16}%
+1991448850n^{14}-1323n^{12}\left(  17018L^{4}-6494305\right)  \right.  \nonumber \\
& +650n^{10}\left(  10847042-687771L^{4}-23310L^{6}\right)  \nonumber \\
& -69n^{8}\left(  112885L^{8}+3576100L^{6}+15774908L^{4}-8000000\right)  \nonumber \\
& -924n^{6}L^{4}\left(  4637L^{6}+106904L^{4}+443260L^{2}+264000\right)  \nonumber \\
& -  \left. 21280n^{4}L^{8}\left(  127L^{4}+1726L^{2}+3900\right)  
-31008n^{2} L^{12}\left(  57L^{2}+350\right)   -930240L^{16} \right),  \\
\varepsilon_{9}(n^{2},L^{2})   & =\frac{n^{16}}{4194304}\left(  408787995n^{18}%
+15278638650n^{16}-341n^{14}\left(  416518L^{4}-289086315\right)  \right.  \nonumber  \\
& -290n^{12}\left(  322478L^{6}+13572891L^{4}-508132290\right)  \nonumber  \\
& -9n^{10}\left(  4892853L^{8}+244792100L^{6}+1786929620L^{4}%
-4297880080\right) \nonumber   \\
& -1300n^{8}L^{4}\left(  16555L^{6}+663164L^{4}+5343868L^{2}+8175108\right)
\nonumber  \\
& -184n^{6}L^{6}\left(  68435L^{6}+1771020L^{4}+10489824L^{2}+12000000\right)
\nonumber  \\
& \left.  -170016n^{4}L^{10}\left(  2160+733L^{2}+49L^{4}\right)
-769120n^{2}L^{14}\left(  7L^{2}+48\right)  -2615008L^{18}\right), \\
\varepsilon_{10}(n^{2},L^{2})   & =-\frac{n^{18}}{33554432}\left(  9724330239n^{20}%
+469170488020n^{18}\right.  \nonumber  \\
& -21n^{16}\left(  175194855L^{4}-203670581987\right)  -1122n^{14}\left(
2123874L^{6}+121493988L^{4}-9238989815\right) \nonumber   \\
& -1023n^{12}\left(  1005969L^{8}+75201300L^{6}+829658295L^{4}%
-5866391024\right) \nonumber   \\
& -116n^{10}\left(  3765438L^{10}+248096745L^{8}+3398870376L^{6}%
+9733541337L^{4}-3239200000\right) \nonumber   \\
& -27n^{8}L^{4}\left(  8514023L^{8}+379043252L^{6}+4451273932L^{4}%
+13108880480L^{2}+6339200000\right) \nonumber   \\
& -2340n^{6}L^{8}\left(  63925L^{6}+1738820L^{4}+13359364L^{2}%
+23240000\right) \nonumber   \\
& \left.  -215280n^{4}L^{12}\left(  483L^{4}+7716L^{2}+28000\right)
-33153120n^{2}L^{16}\left(  2L^{2}+15\right)  -29995680L^{20}\right).
\end{align}

\begin{figure}%
\centering
\includegraphics[width=7cm]{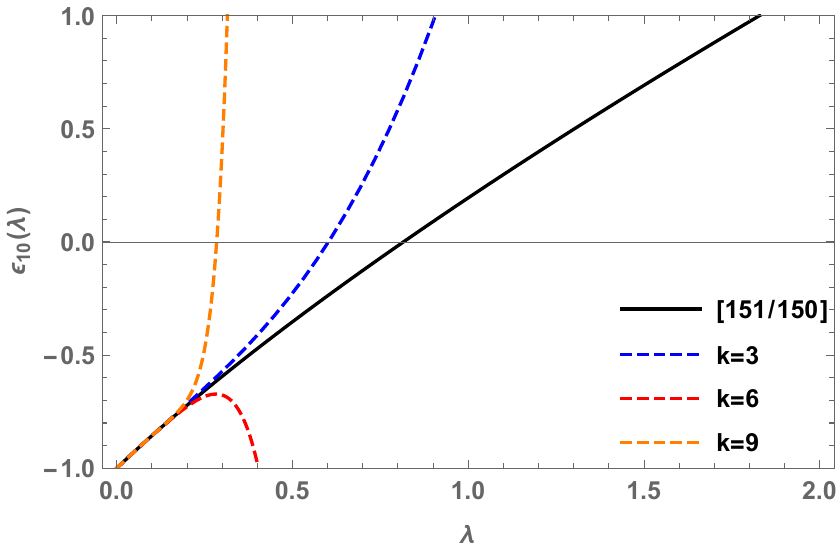}%
\qquad
\includegraphics[width=7cm]{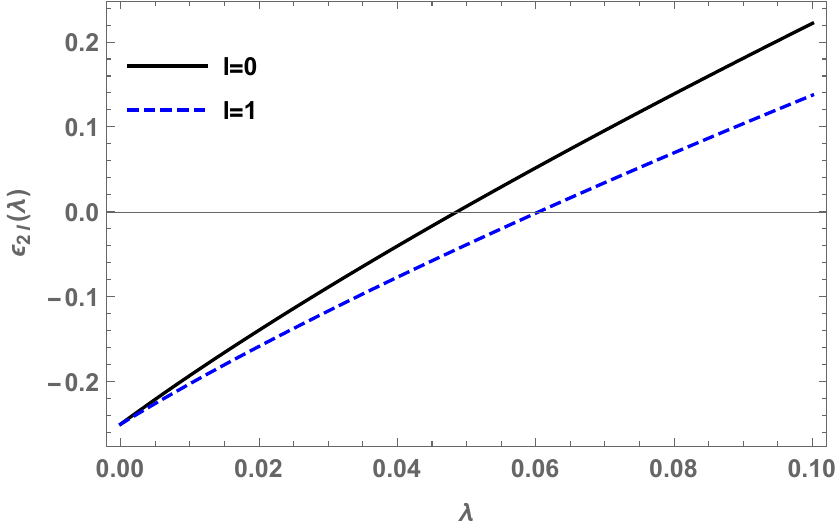}%
\qquad
\includegraphics[width=7cm]{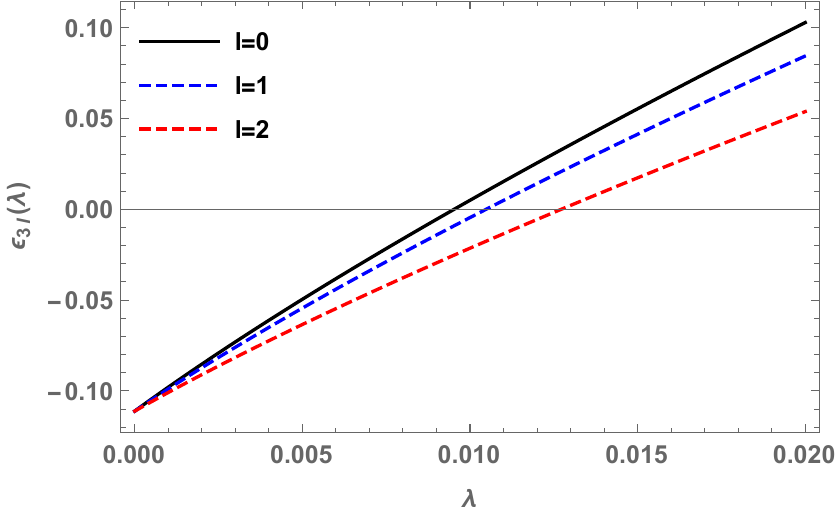}%
\qquad
\includegraphics[width=7cm]{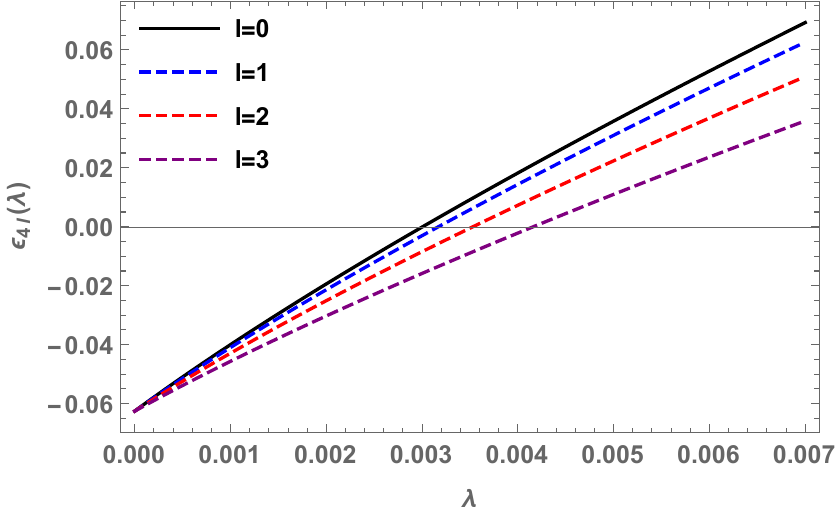}%
\caption{Energy dependence on $\lambda$ of the lowest energy levels of the Cornell potential reconstructed with the 
  $[151/150](\lambda)$ Pad\'{e} approximant. For the ground states we also show results for the series calculated up to ${\cal{O}}(\lambda^{k})$ 
  for $k=3,6,9$. Notice that $\epsilon_{nl}$ {\it{decreases}} with $l$.}
\label{EnlC}
\end{figure}

The dependence on $\lambda$ of the ground state energy is shown in Fig. (\ref{EnlC}) for calculations up to $\lambda^{k}$ with $k=3,6,9$. Clearly,
the series for the energy levels have small convergence radius. However, the $\epsilon_{nl}(\lambda)$ function can be reconstructed from its Taylor 
series using the Pad\'{e} approximants $[M/N](\lambda)$. The actual value of the function lies between the $[N/N](\lambda)$ and 
the $[(N+1)/N](\lambda)$. We can use this result to estimate the uncertainty in the reconstruction which is specially important for large 
values of $\lambda$. Unless stated explicitly, in the following we will use the reconstruction of the energy levels with $N=150$, which requires 
a calculation of the corresponding series to order $\lambda^{301}$. This yields a confident reconstruction of the energy levels, for very large 
values of $\lambda$. The reconstructed energy functions up to $n=4$ are also shown in Fig. (\ref{EnlC}). Notice that, in contrast to screening 
non-confining potentials like the Yukawa and Hulth\'en's potentials, in the case of the Cornell potential, for a given $n$, 
{\emph{ the energy $\epsilon_{nl}(\lambda)$  decreases with $l$}}. This is a surprising result product of the confining linear term in the potential.

A reference point for large values of $\lambda$ is the critical value $\lambda_{cr}$ defined by $\epsilon(\lambda_{cr})=0$. 
Unlike the Yukawa and Hulth\'{e}n potentials, this is not a ionization point because we still have bound states above this critical value 
for the Cornell potential. In Table \ref{Clambdac} we show the results for the critical values of $\lambda$ for the lowest $nl$ levels as 
calculated with $N=5$. The quoted uncertainties correspond to the difference between the $[(N+1)/N]$ and $[N/N]$ Pad\'{e} approximants. 
These uncertainties in the reconstruction are easily reduced using higher $N$, obtaining very precise values for $\lambda_{cr}$, but for the Cornell 
potential these values are not so important thus we quote values with $N=5$ just to have a clear idea of these reference values. 
\begin{table}[!ht]
    \centering
    \begin{tabular}{|c|c|c|}
\hline
   $n$ & $l$ & $\lambda_{cr}$\\
\hline
   1&0&0.8157(7)\\ 
\hline
   2&0&0.048585(5)\\ 
   2&1& 0.06042(2)\\
\hline
   3&0&0.0095239(5)\\ 
   3&1&0.0104780(8)\\ 
   3&2&0.0127192(5)\\ 
\hline
   4&0&0.0030054(1)\\ 
   4&1&0.0031703(1)\\ 
   4&2&0.0035308(3)\\ 
    4&3&0.0041608(5)\\ 
\hline
   5&0&0.00122952(2)\\ 
   5&1&0.00127220(3) \\ 
   5&2&0.00136244(5)\\ 
   5&3&0.0015114(1)\\ 
   5&4&0.0017395(2)\\ 
 \hline
   \end{tabular}
   \qquad
    \centering
    \begin{tabular}{|c|c|c|}
\hline
   $n$ & $l$ & $\lambda_{c}$\\
\hline
       6&0&0.00059256(1)\\ 
   6&1&0.00060674(1)\\ 
   6&2&0.00063623(1)\\ 
   6&3&0.00068347(2)\\ 
   6&4&0.00075267(4)\\
   6&5&0.0008506(1)\\ 
\hline
   7&0&0.000319724(4)\\ 
   7&1&0.000325324(5)\\ 
   7&2&0.000336847(6)\\ 
   7&3&0.00035498(1)\\ 
   7&4&0.00038086(1)\\
   7&5&0.00041625(2)\\ 
   7&6&0.00046379(3)\\ 
\hline
    \end{tabular}
    \qquad
    \centering
     \begin{tabular}{|c|c|c|}
\hline
   $n$ & $l$ & $\lambda_{c}$\\
\hline
8&0&0.000187368(2)\\ 
   8&1&0.000189875(2)\\ 
   8&2&0.000194997(3) \\ 
   8&3&0.000202966(4)\\ 
   8&4&0.000214156(5)\\
   8&5&0.00022912(1)\\ 
   8&6&0.000248667(1)\\ 
   8&7&0.00027394(2)\\ 
\hline
   9&0&0.000116953(1)\\ 
   9&1&0.000118186(1) \\ 
   9&2&0.000120696(2))\\ 
   9&3&0.000124572(2)\\ 
   9&4&0.000129955(4)\\
   9&5&0.000137047(3)\\ 
   9&6&0.000146139(5)\\ 
   9&7&0.00015762(1)\\ 
   9&8&0.00017204(1)\\ 
   \hline
    \end{tabular}
\caption{Critical values of $\lambda$ for the Cornell potential, calculated using the reconstructed functions $\epsilon_{nl}(\lambda)$.}
\label{Clambdac}
\end{table}

\begin{figure}%
\centering
\includegraphics[width=7cm]{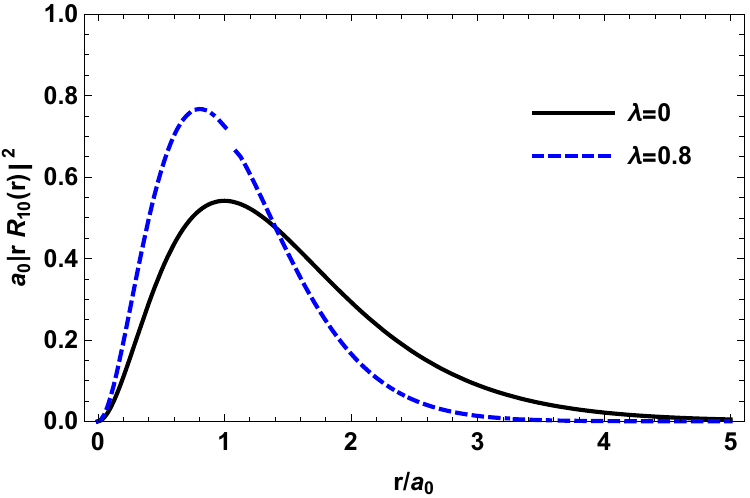} %
\qquad
\includegraphics[width=7cm]{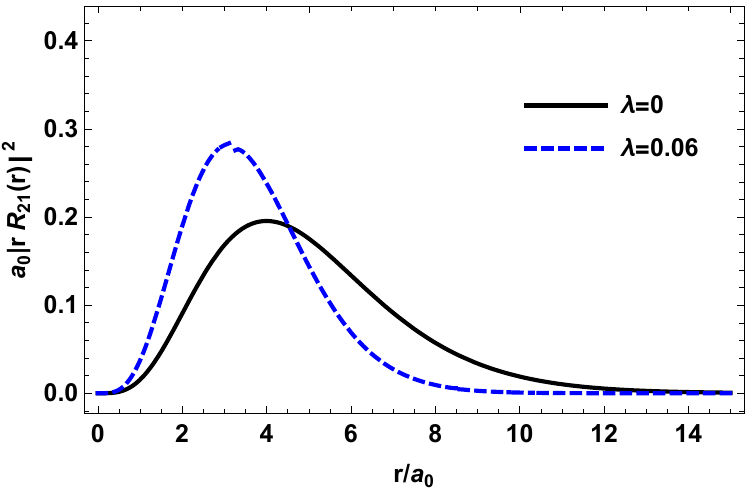} %
\qquad
\includegraphics[width=7cm]{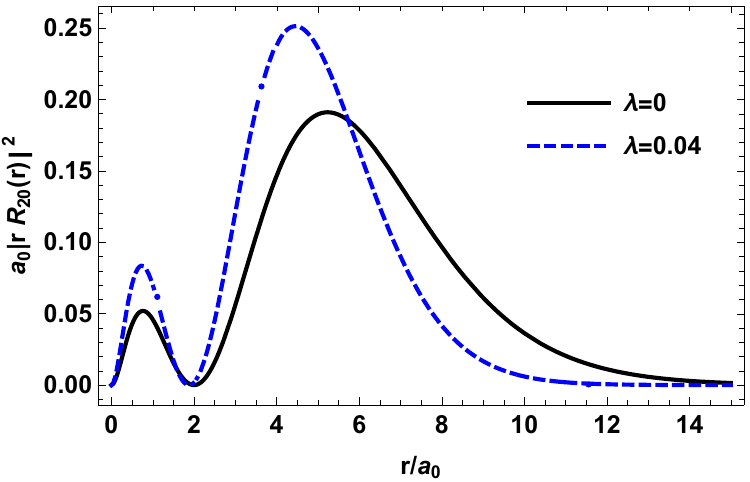} %
\qquad
\includegraphics[width=7cm]{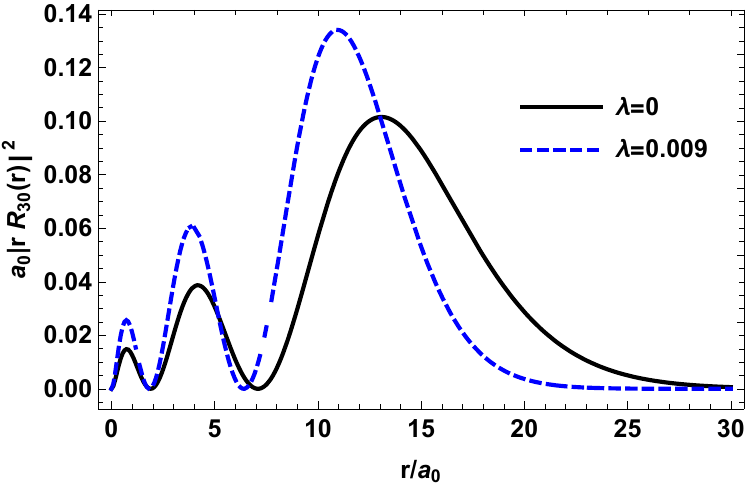} %
\caption{Reconstruction of the probabilities for the lowest lying states of the Cornell potential with the $[5/5](\lambda)$ Pad\'{e} 
  approximant, for values of $\lambda$ close to the critical values for each state and comparison with the Coulomb-like states ($\lambda=0$).  }
 \label{probC}
\end{figure}

Our complete solution allows us to study also the radial probabilities for every state. Similarly to the energies, the eigenstates are 
convergent only for small values of $\lambda$ but they can be reconstructed from the series in $\lambda$ 
using the Pad\'{e} approximants. In this case the reconstruction is more resources demanding.
In Fig. (\ref{probC}) we plot the probabilities for the lowest lying states reconstructed with 
the $[5/5]$ Pad\'{e} approximant, for $\lambda$ close to the critical value in each case, together with the result for the Coulomb-like 
case corresponding to $\lambda=0$. We can see in these plots the second surprising result in the solutions: {\emph{Radial probabilities 
have the same form as the Coulomb-like probabilities, the main peaks grow with $n$, but are shifted to smaller radius by the linear term}}. 
Instead of the conventional delocalization of states produced by 
non-confining screening potential, the confining Cornell potential produces more compact states as we increase the value of $\lambda$. 
It is interesting that the complete solution to the Cornell potential reveals that even for values of $\lambda$ as large as the 
critical values, the shifts are however small and the peaks remain close to the Coulomb-like values. However, in practical 
applications of these results it will be important to asses if the physical values of $\lambda$ are around the critical values or not. 
We will show in the next section that for heavy quarkonium applications, except for the ground state, the 
physical value of $\lambda$ is much larger than its critical value.

\begin{figure}[ht]%
\centering
\includegraphics[width=10cm]{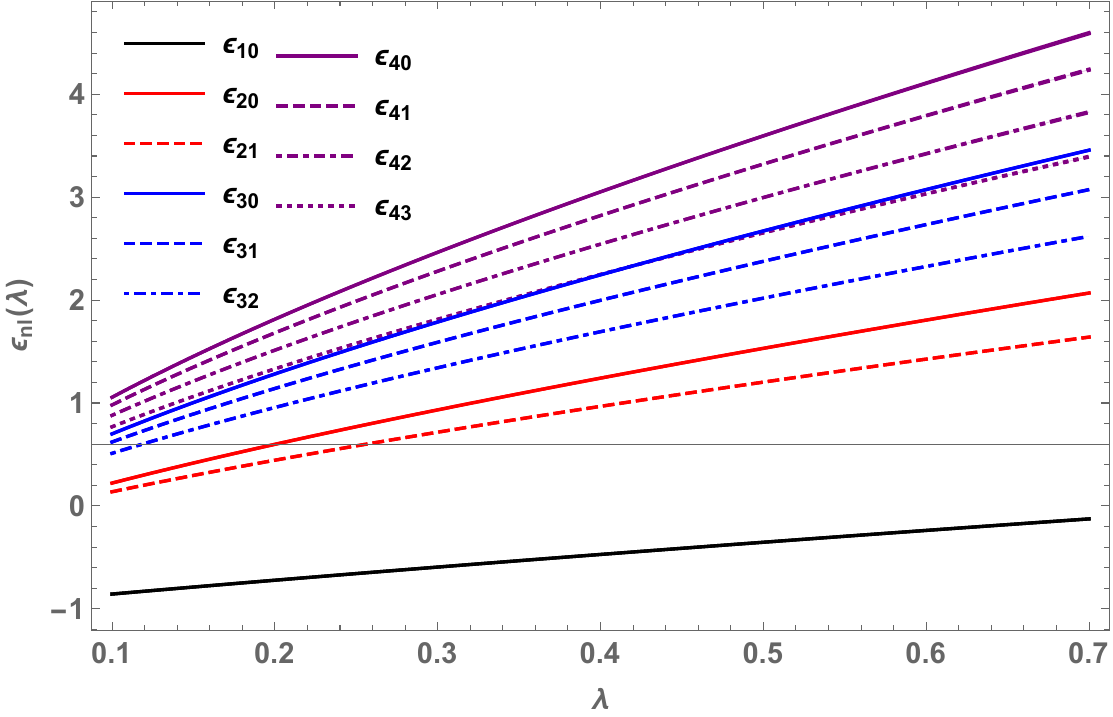}%
\caption{Energy levels, $\epsilon_{nl}(\lambda)$, of the Cornell potential  for $n=1,2,3,4$. For $\lambda \approx 0.4$ the $\epsilon_{43}(\lambda)$ 
  level crosses with the $\epsilon_{30}(\lambda)$ level.  }
\label{Crossing}
\end{figure}

Another interesting result arising in the complete analytical solution of the non-confining screening Yukawa and Hulth\'{e}n potentials 
in Refs. \cite{Napsuciale:2024yrf,Napsuciale:2020ehf,Napsuciale:2021qtw}, is the phenomena of crossing of energy levels. For the Yukawa 
potential this phenomena starts for $n=4$ where, for specific values of the screening parameter close to the critical values, it happens that 
$\epsilon_{43}\ge\epsilon_{50}$. It is interesting to study if a similar effect exists for the Cornell 
potential. The energy levels  for the Cornell potential are shown in Fig. (\ref{Crossing}) for $n=1,2,3,4$. We find that the crossing phenomena 
for the Cornell potential starts with the $\epsilon_{43}(\lambda)$ level which for $\lambda\approx 0.4$ crosses with the $\epsilon_{30}(\lambda)$ . 

\section{Structure of heavy quarkonium from the Cornell potential}
There are many remarkable qualitative and quantitative conclusions that can be drawn on the heavy quarkonium systems from the 
analytical solutions obtained in the previous section. When applied to heavy quarkonium, the solutions for the energy, $E_{nl}(\lambda)$, 
yield the analogous of the Bohr levels of the hydrogen atom. There are striking differences however, 
starting from the fact that non-perturbative strong interactions mimicked by the linear coupling breaks the $l$-degeneracy of the energy 
levels for a given $n$ and produces the inverted spectrum described in the previous section.  

In this section we address the predictions of our complete solution to the Cornell potential for the heavy quarkonium structure. In the following we 
will use the high energy physics customary natural units, i.e., we will set  $\hbar=c=1$.
We start with the typical distance scale associated to the strong coupling constant. 
For heavy quarkonium physics, the parameter $\alpha$ is related to the QCD coupling $\alpha_{s}$ as
\begin{equation}
\alpha=\frac{4}{3}\alpha_{s}.
\label{strongcc}
\end{equation}  
The QCD Bohr radius arising from this coupling constant and denoted in the following 
as $a_{Q}$, is the typical length scale for quarkonium and considering Eq. (\ref{strongcc}) and the fact that for heavy quarkonium the reduced 
mass is $\mu=m_{Q}/2$ we get 
\begin{equation}
a_{Q}=\frac{3}{2}\frac{1}{m_{Q}\alpha_{s}(m_{Q})}.
\label{aQ}
\end{equation}

For heavy quarkonium, the dimensionless coupling $\lambda$ in Eq. (\ref{lambda}) is related to the QCD coupling constant $\alpha_{s}$ and the string 
tension $\sigma$ as
\begin{equation}
\lambda= \frac{27}{8} \frac{\sigma}{m_{Q}^{2}\alpha^{3}_{s} }.
\label{ls}
\end{equation}
The masses of heavy quarkonium states can be written to leading order in a non-relativistic expansion in terms of the eigenvalues of 
the Cornell potential as 
\begin{equation}
M(n^{2S+1}L_{J})(m_{Q},\alpha_{s}(m_{Q}),\lambda_{Q}) =2 m_{Q} + \frac{4}{9}m_{Q} \alpha^{2}_{s}(m_{Q}) \epsilon_{nl}(\lambda_{Q}),
\label{mnl}
\end{equation}
where $m_{Q}, \alpha_{s}(m_{Q})$ and $\lambda_{Q}$ denote the physical values of the corresponding quantities for the heavy quark $Q$.
Notice that for $\lambda=0$ the energy levels reduce to the Coulomb values 
$\epsilon_{nl}(0)=-1/n^{2}$. Our solution allows to go in a controlled manner from the perturbative Coulomb values to the non-perturbative
region mimicked by non-zero values of the normalized string tension $\lambda$.

We remark that we have two well defined scales in Eq. (\ref{mnl})
\begin{align}
\mu_{Q}(m_{Q})&=2m_{Q}, \\
\mu_{B}(m_{Q})&=\frac{4}{9}m_{Q}\alpha^{2}_{s}(m_{Q}).
\end{align}
The scale $\mu_{Q}$ is the scale of the perturbative effects while $\mu_{B}(m_{Q})$ corresponds to the reference energy 
scale for the Bohr-like levels of heavy quarkonium.    

The outstanding result of our complete analytical solution to the Cornell potential is the inverted 
spectrum, i.e. a spectrum where, for a given $n$, states with higher values of $l$ have a lower mass.  This pattern is counter-intuitive 
on the light of results for the non-confining screening potentials like the Yukawa and Hulth\'{e}n potentials 
\cite{Napsuciale:2024yrf,Napsuciale:2020ehf,Napsuciale:2021qtw}, and its comparison against the physical heavy quarkonium 
spectrum is a crucial test for the Cornell potential.
With this aim, we collected from the Review of Particle Properties \cite{ParticleDataGroup:2024cfk}, those states considered as "well established" and 
ordered them in ascending order of the mass values for bottomonium states in Table \ref{bbarbspec}  and for charmonium states in 
Table \ref{cbarcspec} .

\begin{table}[!ht]
    \begin{tabular}{|c|c c c|}
    \hline
   $State$ & $n$ & $^{2S+1}L_J$ & $M_{exp}(GeV)$\\
\hline
    \hline
$\Upsilon(4S)$ & 4& $^{3}S_{1}$ & 10.5794(1)\\ 
 \hline
$\chi_{b2}(3P)$ & 4& $^{3}P_{2}$ & 10.5240(8)\\ 
 \hline       
$\chi_{b1}(3P)$ & 4& $^{3}P_{1}$ & 10.5134(7)\\ 
 \hline     
$\Upsilon(3S)$ & 3& $^{3}S_{1}$ & 10.3551(5)\\ 
 \hline 
$\chi_{b2}(2P)$ & 3& $^{3}P_{2}$ &  10.26865(72)\\ 
 \hline 
$h_{b}(2P)$ & 3& $^{1}P_{1}$ &  10.2598(12)\\ 
 \hline
$\chi_{b1}(2P)$ & 3& $^{3}P_{1}$ & 10.25546(72)\\ 
 \hline 
$\chi_{b0}(2P)$ & 3& $^{3}P_{0}$ &  10.2325(9)\\ 
 \hline
$\Upsilon_{2}(1D)$ & 3& $^{3}D_{2}$ & 10.1637(14)\\ 
 \hline
$\Upsilon(2S)$ & 2& $^{3}S_{1}$ & 10.0234(5)\\ 
 \hline
$\eta_{b}(2S)$& 2 & $^{1}S_{0}$ & 9.999(4)\\ 
 \hline 
 $\chi_{b2}(1P)$ & 2& $^{3}P_{2}$ &  9.91221(57)\\ 
 \hline
$h_{b}(1P)$ & 2& $^{1}P_{1}$ &  9.8993(8)\\ 
 \hline 
$\chi_{b1}(1P)$ & 2& $^{3}P_{1}$ &  9.89278(40)\\ 
 \hline 
 $\chi_{b0}(1P)$ & 2& $^{3}P_{0}$ &  9.85944(73)\\ 
 \hline
 $\Upsilon(1S)$ & 1& $^{3}S_{1}$ & 9.46040(10)\\ 
 \hline
 $\eta_{b}(1S)$& 1 & $^{1}S_{0}$ & 9.3987(20)\\ 
\hline
\end{tabular}
\caption{Experimental results for the spectrum of the bottomonium \cite{ParticleDataGroup:2024cfk}. All the states in this table are considered as 
well established by the Particle Data Group except for the $\eta_{b}(2S)$.}
\label{bbarbspec}
\end{table}
\begin{table}[!ht]
\centering
    \begin{tabular}{|c|c c c|}
    \hline
   $State$ & $n$ & $^{2S+1}L_J$ & $M_{exp}(GeV)$\\
\hline
\hline
    $\psi(4660)^{*}$ & ? & $?$ & $4.641(10)$\\ 
\hline
    $\psi(4415)$ & ? & $^{3}S_{1}$ & $4.415(5)$\\ 
\hline
    $\psi(4360)^{*}$ & ? & $?$ & $4.374(7)$\\ 
 \hline
    $\chi_{c1}(4274)^{*}$ & ? & $?$ & $4.286^{+0.008}_{-0.009}$\\ 
\hline
    $\psi(4230)$ & ? & $^{3}S_{1}$ & $4.2221(23)$\\ 
\hline
    $\psi(4160)$ & ? & $^{3}S_{1}$ & $4.191(5)$\\ 
\hline
    $\chi_{c1}(4140)^{*}$ & ? & $?$ & $4.1465(30)$\\ 
\hline
    $\psi(4040)$ & ? & $^{3}S_{1}$ & $4.040(4)$\\ 
 \hline
    $\chi_{c2}(3930)$& 3 & $^{3}P_{2}$ & $3.9225(10)$\\
\hline
    $\chi_{c0}(3915)^{\dagger}$ & ? & ? & $3.9221(18)$\\ 
\hline
    $\chi_{c1}(3872)^{*}$ & ? & $?$ & $3.87164(6)$\\ 
\hline 
    $\psi(3842)$ & ? & $(^{3}D_{3})$ & $3.84271(20)$\\ 
\hline
    $\psi(3823)$ & ? & $(^{3}D_{2})$ & $3.82351(34)$\\ 
\hline
    $\psi(3770)$ & ? & $(^{3}D_{1})$ & $3.7737(7)$\\ 
\hline
    $\psi(2S)$ & 2 & $^{3}S_{1}$ & $3.686097(11)$\\ 
\hline
    $\eta_{c}(2S)$ & 2 & $^{1}S_{0}$ & $3.6377(9) $\\ 
\hline
    $\chi_{c2}(1P)$ & 2 & $^{3}P_{2}$ & $3.55617(7)$\\ 
\hline
    $h_{c}(1P)$ & 2 & $^{1}P_{1}$ & $3.52537(14)$\\ 
\hline
    $\chi_{c1}(1P)$ & 2 & $^{3}P_{1}$ & $3.51067(5)$\\ 
\hline
    $\chi_{c0}(1P)$ & 2 & $^{3}P_{0}$ & $3.41471(30)$\\ 
\hline
    $J/\psi(1S)$ & 1 & $^{3}S_{1}$ & $3.096900(6)$\\ 
\hline
    $\eta_{c}(1S)$ & 1 & $^{1}S_{0}$ & $2.9841(4)$\\ 
\hline
\end{tabular}
\caption{Charmonium spectrum collected from the Review of Particle Properties \cite{ParticleDataGroup:2024cfk}. We list only those states considered 
as well established by the Particle Data Group. States marked with an asterisk are candidates to be exotic states. States 
marked with $^{\dagger}$ are in doubt in their $J^{P}$ assignment. The $ ^{2S+1}L_{J}$ assignment in parenthesis still need to be confirmed. }
\label{cbarcspec}
\end{table}

We can see in Table \ref{bbarbspec} that for $n=2$, the $l=0$ states, $\Upsilon(2S)$ and $\eta_{b}(2S)$, are heavier than the $l=1$ states, 
$\chi_{b2}(1P),h_{b}(1P),\chi_{b1}(1P),\chi_{b0}(1P) $. Similar results hold for 
$n=3$ where the $l=0$ state, $\Upsilon(3S)$ (the $\eta_{b}(3S)$ state is missing), is heavier than the $l=1$ states, 
$\chi_{b2}(2P),h_{b}(2P),\chi_{b1}(2P),\chi_{b0}(2P) $. 
In turn, these $l=1$ states are heavier than the only $l=2$ state so far discovered in this level, the $\Upsilon_{2}(1D)$ which is a 
$3^{3}D_{2}$ state. The so far discovered bottomonium states in the $n=4$ level follow the same pattern. {\emph{It is clear from the 
mass values in Table \ref{bbarbspec} that the physical spectrum of bottomonium follows the inverted spectrum pattern of the 
Bohr-like levels predicted by the complete analytical solution to the Cornell potential}.

The same pattern is clearly seen for the $n=2$ level of charmonium. Indeed from the experimental mass values in Table \ref{cbarcspec}, 
we can see that in the $n=2$ level, the $l=0$ states $\psi(2s)$ and $\eta_{c}(2S)$ states are heavier than the $l$=1 states 
$\chi_{c2}(1P),h_{c}(1P),\chi_{c1}(1P),\chi_{c0}(1P)$. The identification of the $n^{2S+1}L_{J}$ charmonium states for $n\ge 3$ is still unclear, 
but we can see that a picture consistent with the inverted spectrum is obtained if we consider the $\psi(4040)$ as the $3^{3}S_{1}$ state and 
the $\psi(3842)$, $\psi(3823)$ and $\psi(3770)$ states as the $3^{3}D_{3}$, $3^{3}D_{1}$ and $3^{3}D_{1}$ charmonium states, although it has 
been claimed that the $\psi(3770)$ contain a sizeable tetraquark component \cite{PhysRevD.71.114003}, \cite{BESIII:2022yoo}.

An important physical quantity in heavy quarkonium physics is the size of a given quarkonium configuration. This is important information 
for effective theories of QCD because it is precisely the inverse radius which fixes the scale of the soft modes.  For the Coulomb interaction this 
information is usually obtained from the mean value of the inverse radius. With our complete analytical 
solution we can calculate the mean value of the inverse radius normalized to the strong Bohr radius for every $n,l$ heavy quarkonium configuration as
\begin{equation}
a_{Q}r^{-1}_{nl}(\lambda)= \langle nl | \frac{a_{Q}}{r} |nl\rangle= a_{Q} \int_{0}^{\infty}dr r |R_{nl}(r,\lambda)|^{2}\equiv f^{(-1)}_{nl}(\lambda).
\label{rnl}
\end{equation}
We remark that the mean value of the inverse radius normalized by the strong Bohr radius is a function that depends only on the normalized 
string tension $\lambda$. In Fig. (\ref{rnlfig}) we plot $f^{(-1)}(\lambda)$ for the lowest lying heavy quarkonium $nl$ states, calculated with the 
$[101/100]$ Pad\'{e} approximant. We can see in this plot 
that $r^{-1}_{nl}(\lambda)$ goes from its $l$-independent Coulomb value $r^{-1}_{nl}(0)=1/n^{2}a_{Q}$, to a $l$-dependent value for 
the physical $\lambda_{Q}$, which increases with increasing $\lambda$. We conclude from this plot that the effective radius of heavy quarkonium 
states grows with $n$ but this growing is significantly lower than the Coulomb $n^{2} a_{Q}$ behavior, i.e. heavy quarkonium systems are 
more compact than Coulomb-like systems given by the perturbative term. 

\begin{figure}%
\centering
\includegraphics[width=7cm]{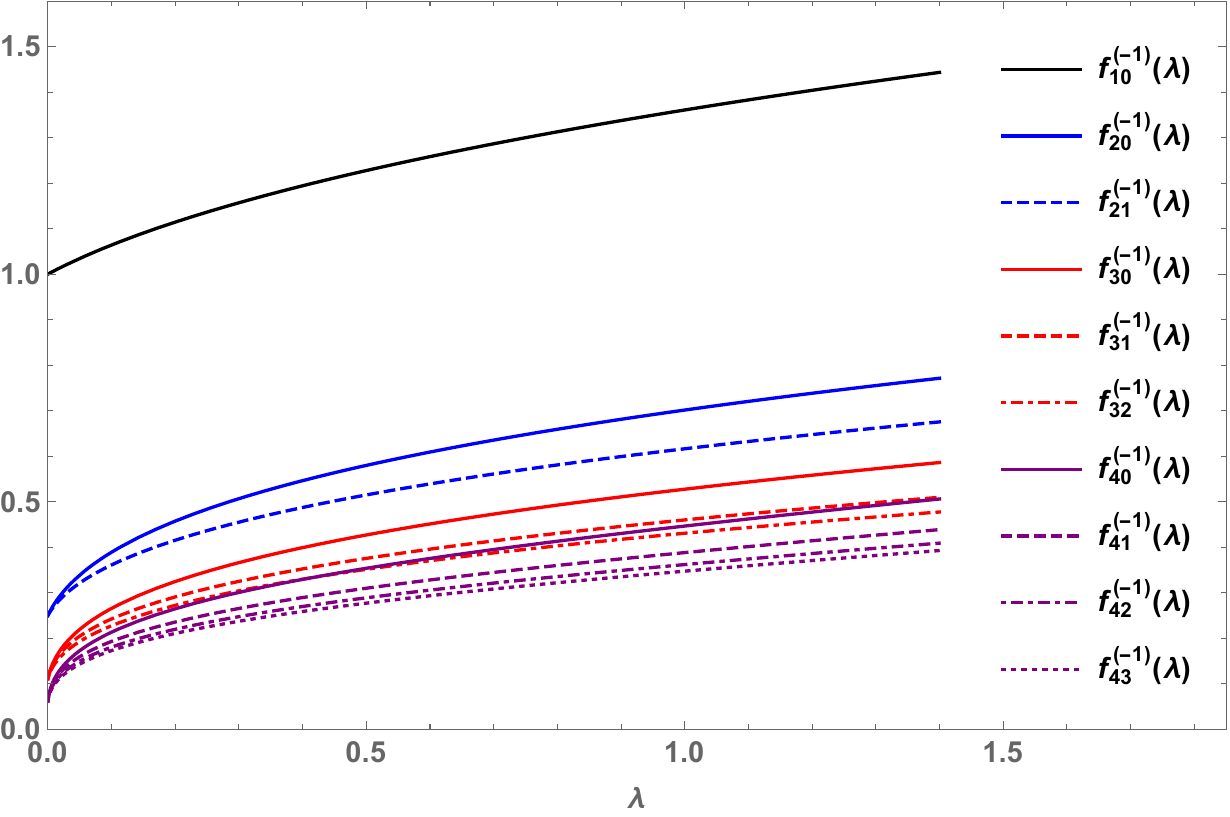} %
\caption{Heavy quarkonium inverse mean radius in units of the Bohr radius, $\langle nl |(\frac{r}{a_{Q}})^{-1}|nl\rangle$, for the lowest lying states.}
 \label{rnlfig}
\end{figure}

\subsection{Quantitative Bohr-like heavy quarkonium systems} 

The qualitative picture of quarkonium structure agrees with the measured heavy quarkonium spectrum and it would be interesting to do a more
quantitative analysis of the predictions in Eq. (\ref{mnl}). A first estimate 
of the physical values of the parameters and involved quantities can be obtained considering that the Borh-like levels $\epsilon_{nl}(\lambda)$ 
correspond to the average values of all the allowed $m$ values in this level. We expect relativistic corrections to be important for real 
heavy quarkonium systems thus the estimates in the remaining of this subsection must be considered only as a starting point toward the 
description of real heavy quarkonium systems. 

For a given $n$, the average values of the masses of the $l=0,1$ states are given by
\begin{align}
\bar{M}_{n0}(\bar{\lambda}_{Q})&= \frac{1}{4}\left( M[n^{1}S_{0}] + 3M[n^{3}S_{1}] \right), \\
\bar{M}_{n1}(\bar{\lambda}_{Q})&= \frac{1}{12}\left( M[n^{3}P_{0}] + 3M[n^{3}P_{1} ]+ 5 M[n^{3}P_{2} ]+3 M[n^{1}P_{1}]  \right),
\end{align}
where hereafter we use a bar for the physical quantities extracted from the physical average masses. 

From the values for physical quarkonium in Tables \ref{bbarbspec},\ref{cbarcspec}, we can see that we have complete experimental 
data only for the $n=1$ and $n=2$ levels in both quarkonium sectors, thus we will use this data.
We get the following average central values for the $n=1,2$ bottomonium and charmonium levels
\begin{align}
\bar{M}_{10}(\bar{\lambda}_{b})&=9444.98~MeV, \quad &\bar{M}_{10}(\bar{\lambda}_{c})=  3068.70 ~MeV,\label{Mbar10} \\
\bar{M}_{20}(\bar{\lambda}_{b})&= 10017.30 ~MeV, \quad &\bar{M}_{20}(\bar{\lambda}_{c})= 3674.00 ~MeV,\label{Mbar20}\\
\bar{M}_{21}(\bar{\lambda}_{b})&= 9899.73 ~MeV, \quad &\bar{M}_{21}(\bar{\lambda}_{c})= 3525.31  ~MeV.\label{Mbar21}
\end{align} 
It is convenient for the purposes of this subsection to rewrite Eq. (\ref{mnl}) as
\begin{align}
\bar{M}_{nl}(\bar{\lambda}_{Q})&= \mu_{Q} (\bar{m}_{Q}) + \mu_{B}(\bar{m}_{Q})\epsilon_{nl}(\bar{\lambda}_{Q}).
\label{mnlav}
\end{align}
The Bohr level values of $\bar{\lambda}_{Q}$, can be obtained from these relations for $n=1,2$ as the solution to
\begin{align}
\frac{\epsilon_{20}(\bar{\lambda}_{Q})- \epsilon_{10}(\bar{\lambda}_{Q})}{\epsilon_{20}(\bar{\lambda}_{Q})- \epsilon_{21}(\bar{\lambda}_{Q})}=
\frac{\bar{M}_{20}(\bar{\lambda}_{Q})- \bar{M}_{10}(\bar{\lambda}_{Q})}{\bar{M}_{20}(\bar{\lambda}_{Q})- \bar{M}_{21}(\bar{\lambda}_{Q})}.
\label{lambdabareq}
\end{align}
The experimental values for the ratio on the right hand side are obtained from the average values in Eqs. (\ref{Mbar10},\ref{Mbar20},\ref{Mbar21}) as 
\begin{align}
\frac{\bar{M}_{20}(\bar{\lambda}_{b})- \bar{M}_{10}(\bar{\lambda}_{b})}{\bar{M}_{20}(\bar{\lambda}_{b})- \bar{M}_{21}(\bar{\lambda}_{b})}=4.868\pm 0.037, \qquad
\frac{\bar{M}_{20}(\bar{\lambda}_{c})- \bar{M}_{10}(\bar{\lambda}_{c})}{\bar{M}_{20}(\bar{\lambda}_{c})- \bar{M}_{21}(\bar{\lambda}_{c})}=4.071\pm 0.005. 
\end{align}
Solving equation (\ref{lambdabareq}) for bottomonium and charmonium, we get the following estimates of the 
central values of the normalized string tension, extracted from the Bohr-like (purely non-relativistic) heavy quarkonium levels 
\begin{align}
\bar{\lambda}_{b}=0.884, \qquad
\bar{\lambda}_{c}=1.694.
\label{Bohrlambda}
\end{align}
Although these values are natural in the sense that for dimensionless parameters we expect ${\cal{O}}(1)$ values, we must consider them as gross 
estimates of the physical values, until we have clear idea on the size of the relativistic corrections. The lack of estimates of the size of these corrections 
prevent a serious calculation of the uncertainties in these quantities at this point, this is the reason why we refrain from quoting uncertainties in these quantities. 
Meanwhile, pushing forward this purely non-relativistic 
quantitative picture, we can now obtain from the same data the  gross values of the scales $\mu_{Q} (\bar{m}_{Q})$ and $\mu_{B}(\bar{m}_{Q})$, which 
are given by
\begin{align}
\mu_{B} (\bar{m}_{Q})&=\frac{\bar{M}_{20}(\bar{\lambda}_{Q})-\bar{M}_{10}(\bar{\lambda}_{Q})}{\epsilon_{20}(\bar{\lambda}_{Q})- \epsilon_{10}(\bar{\lambda}_{Q})}, \\
\mu_{Q}(\bar{m}_{Q})&= \frac{\epsilon_{20}(\bar{\lambda}_{Q})\bar{M}_{10}(\bar{\lambda}_{Q}) - \epsilon_{10}(\bar{\lambda}_{Q})\bar{M}_{20}(\bar{\lambda}_{Q})}%
{\epsilon_{20}(\bar{\lambda}_{Q})- \epsilon_{10}(\bar{\lambda}_{Q})}.
\end{align}
Our solutions yield the following values for the energy levels  
\begin{align}
\epsilon_{10}(\bar{\lambda}_{b})&= 0.073, \quad &\epsilon_{10}(\bar{\lambda}_{c})= 0.875, \\
\epsilon_{20}(\bar{\lambda}_{b})&= 2.527 , \quad &\epsilon_{20}(\bar{\lambda}_{c})= 4.299, \\
\epsilon_{21}(\bar{\lambda}_{b})&= 2.014, \quad &\epsilon_{21}(\bar{\lambda}_{c})= 3.458 .
\end{align}
The values  obtained for the energy scales at the Bohr values of the normalized string tension are
\begin{align}
\mu_{B} (\bar{m}_{b})&=233.25~ MeV, \quad &\mu_{B} (\bar{m}_{c})=176.80 ~ MeV, \\
\mu_{Q} (\bar{m}_{b})&=9427.92 ~ MeV, \quad &\mu_{Q} (\bar{m}_{c})= 2914.00~ MeV. 
\end{align}
These values for the physical scales yields the following values for the strong coupling constant and heavy quark masses
\begin{align}
\bar{m}_{b}&=4713.96~MeV, \quad &\alpha_{s}(\bar{m}_{b})= 0.3337, \\
\bar{m}_{c}&=1457.0 ~MeV, \quad &\alpha_{s}(\bar{m}_{c})= 0.5225.
\end{align}
We insist in that these values are obtained in an purely non-relativistic calculation and should be considered only as a first approximation to the 
real values.  We can have an idea of the variations using instead the data on the masses of the $1^{3}S_{1}$, $2^{3}S_{1}$, and $3^{3}S_{1}$, 
for the bottomonium states to fix $\bar{\lambda}_{b}$, which yields $\bar{\lambda}_{b}\simeq 0.4$ and similar values for $\alpha_{s}(\bar{m}_{b})$ and 
$\mu_{Q}(\bar{m}_{b}),\mu_{B}(\bar{m}_{b})$. This procedure, however, would be inconsistent because the Bohr-like levels $\epsilon_{n0}$ must correspond 
to the average of the $n^{3}S_{1}$ and $n^{1}S_{0}$, but it exhibits that the extraction of the physical values of $\lambda_{Q}$ may be sensible to corrections 
to the Cornell potential. 

Once fixed the Bohr values of $\bar{\lambda}_{Q}$ and of the energy scales, our solution yields the following values 
for the energy of the $n=3,4$ excited states
\begin{align}
\epsilon_{30}(\bar{\lambda}_{b})= 4.121, \quad & \epsilon_{30}(\bar{\lambda}_{c})= 6.666, \\
\epsilon_{31}(\bar{\lambda}_{b})= 3.663, \quad & \epsilon_{31}(\bar{\lambda}_{c})= 5.919, \\
\epsilon_{32}(\bar{\lambda}_{b})= 3.128, \quad & \epsilon_{32}(\bar{\lambda}_{c})= 5.076, \\
\epsilon_{40}(\bar{\lambda}_{b})= 5.448, \quad & \epsilon_{40}(\bar{\lambda}_{c})= 8.762 , \\
\epsilon_{41}(\bar{\lambda}_{b})= 5.024, \quad & \epsilon_{41}(\bar{\lambda}_{c})= 8.067, \\
\epsilon_{42}(\bar{\lambda}_{b})= 4.537 , \quad & \epsilon_{42}(\bar{\lambda}_{c})= 7.292, \\
\epsilon_{43}(\bar{\lambda}_{b})= 4.028 , \quad & \epsilon_{43}(\bar{\lambda}_{c})= 6.488, 
\end{align}
and from Eq. (\ref{mnlav}) the following predictions of the Cornell potential for the average masses of the $n=3,4$ excited states are obtained
\begin{align}
\bar{M}_{30}(\bar{\lambda}_{b})&= 10389.10  ~MeV, \quad & \bar{M}_{30}(\bar{\lambda}_{c})&= 4092.46~MeV, \\
\bar{M}_{31}(\bar{\lambda}_{b})&= 10282.20  ~MeV, \quad & \bar{M}_{31}(\bar{\lambda}_{c})&= 3960.52 ~MeV, \label{Mbar31}\\
\bar{M}_{32}(\bar{\lambda}_{b})&= 10157.40 ~MeV, \quad & \bar{M}_{32}(\bar{\lambda}_{c})&= 3811.48~MeV, \\
\bar{M}_{40}(\bar{\lambda}_{b})&= 10698.70~MeV, \quad & \bar{M}_{40}(\bar{\lambda}_{c})&= 4463.19  ~MeV, \\
\bar{M}_{41}(\bar{\lambda}_{b})&= 19599.80 ~MeV, \quad & \bar{M}_{41}(\bar{\lambda}_{c})&=4340.29  ~MeV, \\
\bar{M}_{42}(\bar{\lambda}_{b})&= 10486.20  ~MeV, \quad & \bar{M}_{42}(\bar{\lambda}_{c})&= 4203.22 ~MeV, \\
\bar{M}_{43}(\bar{\lambda}_{b})&= 10367.40  ~MeV, \quad & \bar{M}_{43}(\bar{\lambda}_{c})&= 4061.04 ~MeV. 
\end{align}

From data in Table \ref{bbarbspec}, the only available average value beyond those in Eqs. (\ref{Mbar10},\ref{Mbar20},\ref{Mbar21}) used to fix the free 
parameters, is the average mass of the $n=3$, $l=1$ level, where we have data on the masses of all the members of the corresponding multiplet. Other 
multiplets have missing states. From data  in Table \ref{bbarbspec} we obtain $\bar{M}_{31}(\bar{\lambda}_{b})_{exp}= 10260.10$. This central value is 
$0.2\%$ off the central value prediction in Eq. (\ref{Mbar31}). This correspond to a difference of the order of $22~ MeV$ which should be the order of 
magnitude of the corrections to the Cornell potential, which include relativistic corrections. In this concern, we can estimate from our solutions the 
average $v^{2}$ for the different heavy quarkonium configurations, which can give us an idea of the size of 
relativistic corrections. This observable is given by
\begin{equation}
\langle nl | v^{2} |nl \rangle=\frac{4}{m_{Q}}\langle nl | H_{0}+ \frac{4\alpha_{s}}{3r} - \sigma r |nl \rangle 
= \alpha^{2}_{s} C_{nl}(\lambda),
\label{v2nl}
\end{equation}
with
\begin{equation}
C_{nl}(\lambda)=\frac{16}{9} \left( \epsilon_{nl}(\lambda) +  2 f^{(-1)}_{nl}(\lambda) - \lambda f^{(1)}_{nl}(\lambda) \right),
\end{equation}
where
\begin{equation}
 f^{(k)}_{nl}(\lambda)\equiv \langle nl | \left(\frac{r}{a_{Q}}\right)^{k}|nl\rangle= a^{-k}_{Q} \int_{0}^{\infty}dr r^{2+k} |R_{nl}(r)|^{2}.
 \label{fknl}
\end{equation}
are dimensionless functions depending only on $\lambda$.

Using the virial theorem it can be shown that for the Cornell potential
\begin{equation}
\epsilon_{nl}(\lambda)=-f^{-1)}_{nl}(\lambda) + \frac{3}{2} \lambda f^{(1)}_{nl}(\lambda), 
\end{equation}
and the proportionality constant can also be written as
\begin{equation}
C_{nl}(\lambda)=\frac{8}{9} \left( 2 f^{(-1)}_{nl}(\lambda) + \lambda f^{(1)}_{nl}(\lambda) \right).
\label{Cnleq}
\end{equation}

Equations (\ref{v2nl},\ref{Cnleq}) show that the average value of $v^{2}$ is indeed proportional of $\alpha^{2}_{s}$ as expected from the 
NRQCD counting rules. However, the proportionality constant depends in a complicated manner on the string tension. The detail of this 
dependence can be obtained from our complete solution to the Cornell potential and it is shown in Fig. (\ref{Cnl}) where we plot the proportionality 
constant $C_{nl}$ as a function of $\lambda$ for the lowest lying $n=1,2,3,4$ states, calculated with the $[101/100]$ Pad\'{e} approximant. 
We can see in this plot that the proportionality constant changes its behavior from its Coulombian value $C_{nl}(0)=\frac{16}{9n^{2}}$ 
decreasing with $n$ and $l$-independent, to a value around $2$  which is $l$-dependent, for $\lambda\approx 0.4$. {\it{The general lesson from 
this plot is that relativistic corrections will be more important than expected for both ground and excited heavy quarkonium states}}, although a 
precise quantification of this effect must await for a more 
precise extraction of the physical value $\lambda_{Q}$ from data.  We devote the next section to the calculation of the leading 
relativistic spin-dependent corrections which in addition to 
modify the predictions for the Bohr levels break the $m$-degeneracy in $\epsilon_{nl}(\lambda)$ through the spin-dependent interactions.
In preparation, we notice that there is an obvious mass hierarchy in the experimental fine splittings of heavy quarkonium which can be seen 
in Tables \ref{bbarbspec}, \ref{cbarcspec}. Indeed, from the measured values of the masses of the $n=1$ and $n=2$ levels of heavy quarkonium, 
systematically for the $s$-waves, the $n^{3}S_{1}$ state is heavier that the $n^{1}S_{0}$ state. Also, for the $n=2$ and $n=3$ 
(in the bottomonium case) $p$-wave states, the masses are such that systematically 
$M_{\chi_{b2}(nP)} > M_{h_{b}(nP)}>M_{\chi_{b1}(nP)}>M_{\chi_{b0}(nP)} $.

\begin{figure}%
\centering
\includegraphics[width=7cm]{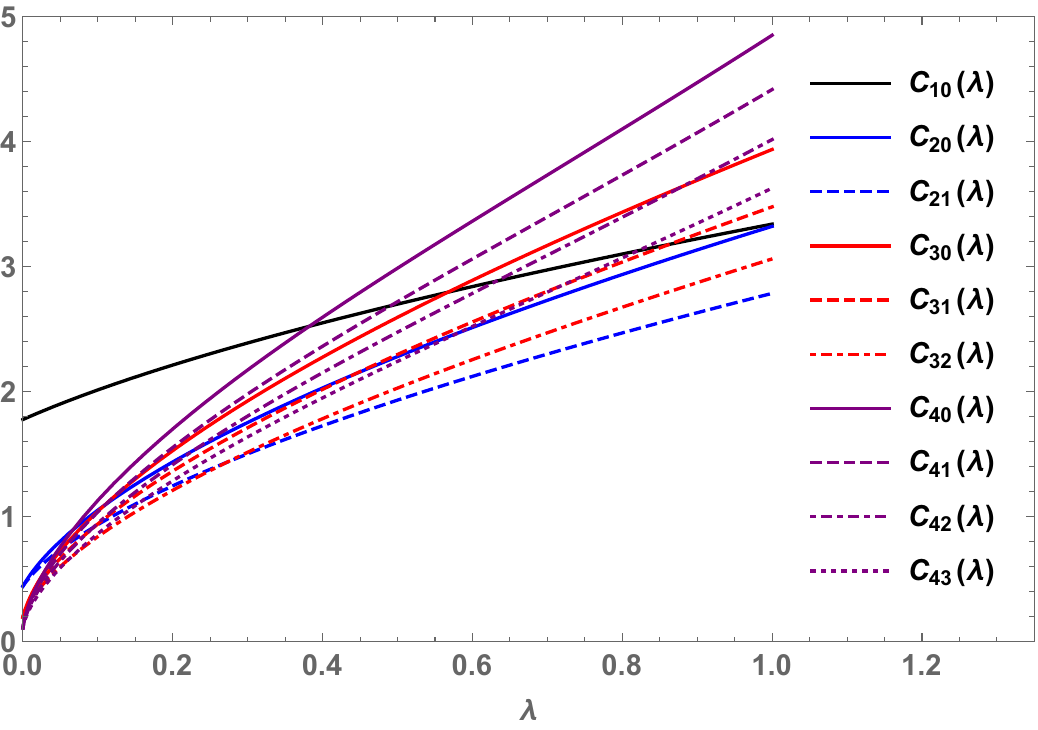} %
\caption{Proportionality constant $C_{nl}(\lambda)$ between $\langle nl |v^{2}|nl\rangle$ and $\alpha^{2}_{s}$ as a function of $\lambda$ for the lowest lying states.}
 \label{Cnl}
\end{figure}

\section{Towards real heavy quarkonium: spin-dependent relativistic corrections}

Relativistic corrections are expected from Eq. (\ref{v2nl}) to be of order $\alpha_{s}^{4}$ and to yield a closer approach to real
heavy quarkonium systems. The derivation of the leading relativistic corrections has been done using diverse methods
\cite{Eichten:1979pu,Eichten:1980mw,Buchmuller:1981fr,Gromes:1983pm, Gromes:1984ma, Falkensteiner:1984su,Barchielli:1986zs, 
Barchielli:1988zp,Lucha:1991vn} and the corresponding operators can be classified into spin-independent and spin-dependent interactions.
The leading spin-independent corrections include conventional corrections to the kinetic energy and velocity-dependent terms  
\cite{Barchielli:1986zs, Barchielli:1988zp}, which shifts the Bohr-like energy levels $\epsilon_{nl}(\lambda)$ but do not remove the 
$m$-degeneracy of the $l$ orbitals for a fixed $n$. It is not the aim of this work to do a complete phenomenological analysis of the 
relativistic corrections and we will focus only on the fine splittings in heavy quarkonium produced by the spin-dependent interactions.
However, there are interesting conclusions arising from this analysis concerning the $1/m_{Q}$ expansion and we will consider 
explicitly the corrections to the kinetic energy in the spin-independent sector to illustrate the point. These corrections are induced by
\begin{align}
H_{K}&= - \frac{p^{4}}{4m^{3}_{Q}} =  -\frac{1 }{4m_{Q} }\left(H_{0} 
+ \frac{4}{3}\frac{\alpha_{s}}{r} - \frac{8}{27} m^{2}_{Q}\alpha^{3}_{s} \lambda r\right)^{2}.
\end{align} 
The corrections to $ E_{nl}$ due to this term, as calculated in perturbation theory are given by
\begin{align}
\Delta E^{si}_{nl}&=
 - \frac{4}{81} m_{Q}\alpha^{4}_{s}\left( 
\epsilon^{2}_{nl}(\lambda) + 4 f^{(-2)}_{nl}(\lambda) +  \lambda^{2} f^{(2)}_{nl}(\lambda) + 4 \epsilon_{nl} (\lambda) f^{(-1)}_{nl}(\lambda) 
- 2 \lambda \epsilon_{nl} (\lambda) f^{(1)}_{nl}(\lambda) -   4 \lambda \right).
\label{deltaesi}
\end{align}

\subsection{Spin-dependent interactions}

The proper description of the physical spectrum of heavy quarkonium requires to consider the spin-dependent 
interactions responsible for the splitting of the Bohr energy levels $\epsilon_{nl}(\lambda)$. These interactions arise from the leading relativistic 
corrections and for the Cornell potential are given by 
\begin{equation}
H_{SD}= \frac{8 \alpha_{s}}{9m^{2}_{Q}} \bm{S}_{1}\cdot \bm{S}_{2} \frac{\delta(r)}{r^{2}} 
+ \frac{2\alpha_{s}}{m^{2}_{Q}}  \frac{\bm{L}\cdot \bm{S}}{r^{3}} 
+ \frac{\alpha_{s}}{3m^{2}_{Q}}  \frac{S_{12}}{r^{3}}
- \frac{\sigma}{2m^{2}_{Q} }\frac{\bm{L}\cdot \bm{S}}{r},
\label{sdint}
\end{equation} 
where 
\begin{equation}
S_{12}=4[3(\bm{S}_{1}\cdot \hat{\bm{r}})(\bm{S}_{2}\cdot \hat{\bm{r}}) - \bm{S}_{1}\cdot \bm{S}_{2}].
\end{equation}
The leading relativistic corrections have a $1/m^{2}_{Q}$ suppression and are expected to be small.
The first term in Eq. (\ref{sdint}) is the analogous of the hyperfine splitting in atomic physics where it is suppressed due to the small electron 
to proton mass ratio. Here, this term is of the same order as the remaining fine structure terms and we will denote all terms in Eq. (\ref{sdint}) as fine 
structure interactions. We can treat these interactions perturbatively due to their formal $1/m^{2}_{Q}$ suppression. To leading order we need to calculate 
the spin-dependent matrix elements
\begin{equation}
\langle\bar{Q}Q[n^{2S+1}L_{J}]|O_{i}|\bar{Q}Q[n^{2S+1}L_{J}]\rangle,
\end{equation}
for $O_{1}=\bm{S}_{1}\cdot \bm{S}_{2} \frac{\delta(r)}{r^{2}} $, $O_{2}=\frac{\bm{L}\cdot \bm{S}}{r^{3}} $, $O_{3}= \frac{S_{12}}{r^{3}}$ and 
$O_{4}=\frac{\bm{L}\cdot \bm{S}}{r} $. 
The angular momentum part yields the well defined $n$-independent factors
\begin{align}
\langle  \bm{S}_{1}\cdot \bm{S}_{2}\rangle&=\frac{1}{2}\left[s(s+1)-\frac{3}{2}\right], \\
\langle \bm{L}\cdot \bm{S}\rangle&= \frac{1}{2}[j(j+1)-l(l+1)-s(s+1)], \\
\langle S_{12} \rangle&=
\frac{2[2l(l+1)s(s+1)- 3 \langle \bm{L}\cdot \bm{S}\rangle- 6\langle \bm{L}\cdot \bm{S}\rangle^{2}]}{(2l-1)(2l+3)}.
\end{align}

The explicit spin factors for the lowest lying angular momentum configurations are given in Table \ref{spinfac}.
\begin{table}[!ht]
\centering
    \begin{tabular}{|c|c|c|c|c|c|c|c|c|c|c|}
    \hline
   Operator& $^{1}S_0$ & $^{3}S_1$ & $^{1}P_1$  & $^{3}P_0$  & $^{3}P_1$ & $^{3}P_2$ & $^{1}D_2$& $^{3}D_1$& $^{3}D_2$& $^{3}D_3$ \\
\hline
$\langle \bm{S}_{1}\cdot \bm{S}_{2}\rangle$ & $-\frac{3}{4}$ & $\frac{1}{4}$ & $-\frac{3}{4}$ & $\frac{1}{4}$ & $\frac{1}{4}$ & $\frac{1}{4}$ & 
 $-\frac{3}{4}$ & $\frac{1}{4}$ & $\frac{1}{4}$ & $\frac{1}{4}$ \\
$\langle \bm{L}\cdot \bm{S}\rangle$ & $0$ & $0$ & $0$ & $-2$ & $-1$ & $1$ & $0$ & $-3$ & $-1$ & $2$ \\
$\langle S_{12}\rangle$ & $0$ & $0$ & $0$ & $-4$ & $2$ & $-\frac{2}{5}$ & $0$ & $-2$ & $2$ & $-\frac{4}{7}$\\
\hline
\end{tabular}
\caption{Spin factors for the lowest lying angular momentum configurations.}
\label{spinfac}
\end{table}

The corrections to $E_{nl}$ due to the spin-dependent term to the mass of heavy quarkonium, as calculated in perturbation theory, are given by
\begin{align}
\Delta E^{sd}_{nl}(\lambda)&=
 \frac{8\alpha_{s}}{9m^{2}_{Q}} \langle \bm{S}_{1}\cdot \bm{S}_{2}\rangle \left\langle \frac{\delta (r)}{r^{2}}\right\rangle_{nl} \nonumber\\
&+ \left(\frac{2\alpha_{s}}{m^{2}_{Q}} \left\langle\frac{1}{r^{3} }\right\rangle_{nl} 
- \frac{\sigma}{2m^{2}_{Q}} \left\langle\frac{1}{r}\right\rangle_{nl} \right) \langle \bm{L}\cdot \bm{S}\rangle 
+\frac{\alpha_{s}}{3m^{2}_{Q}}\langle S_{12}\rangle \left\langle\frac{1}{r^{3}} \right\rangle_{nl}.
\label{mnslj}
\end{align}
This result can be written as
\begin{align}
\Delta E_{nl}^{sd}(\lambda)= \frac{16}{243} m_{Q}\alpha^{4}_{s}   
 \left[ f^{ss}_{nl}(\lambda) 4 \langle \bm{S}_{1}\cdot \bm{S}_{2}\rangle
+ \left(9 f^{(-3)}_{nl} (\lambda)- \frac{3}{2} \lambda f^{(-1)}_{nl} (\lambda) \right)  \langle \bm{L}\cdot \bm{S}\rangle 
+\frac{3}{2} f^{(-3)}_{nl}(\lambda) \langle S_{12}\rangle \right],
\label{deltaesd}
\end{align}
where 
\begin{equation}
 f^{ss}_{nl}(\lambda)=a^{3}_{Q} \int_{0}^{\infty}dr ~\delta(r)|R_{nl}(r,\lambda)|^{2}= a^{3}_{Q} |R_{nl}(0,\lambda)|^{2}
 \label{fssnl}
\end{equation}
and $f^{(k)}(\lambda)$ are given in Eq. (\ref{fknl}).

\subsection{Heavy quarkonium spectrum to order $\alpha_{s}^{4}$}

Considering both spin-independent and spin-dependent interactions, the mass of heavy quarkonium in a $n^{2S+1}L_{J}$ configuration 
to order $\alpha^{4}_{s}$ is given by 
\begin{align}
M[n^{2S+1}L_{J}] &= 2m_{Q} + \frac{4}{9}m_{Q}\alpha^{2}_{s} \epsilon_{nl}(\lambda)  
- \frac{64}{243} m_{Q}\alpha^{4}_{s} \Delta^{SI}_{nl}(\lambda)\nonumber \\
&+\frac{64}{243} m_{Q}\alpha^{4}_{s}   \left[ f^{ss}_{nl}(\lambda)  \langle \bm{S}_{1}\cdot \bm{S}_{2}\rangle
+ \left(\frac{9}{4} f^{(-3)}_{nl} (\lambda)- \frac{3}{8} \lambda f^{(-1)}_{nl} (\lambda) \right)  \langle \bm{L}\cdot \bm{S}\rangle 
+\frac{3}{8} f^{(-3)}_{nl}(\lambda) \langle S_{12}\rangle \right].
\label{mnslj}
\end{align}
with
\begin{equation}
 \Delta^{SI}_{nl}(\lambda)= \frac{3}{16} \epsilon^{2}_{nl}(\lambda) + \frac{3}{4} f^{(-2)}_{nl}(\lambda) +  \frac{3}{16} \lambda^{2} f^{(2)}_{nl}(\lambda) 
 + \frac{3}{8} \epsilon_{nl}(\lambda) \left(2f^{(-1)}_{nl}(\lambda) 
-  \lambda f^{(1)}_{nl}(\lambda)\right) -   \frac{3}{4} \lambda.
\end{equation}
This is the second main result of this paper and there are several worth remarks on Eq. (\ref{mnslj}). 
Firstly, notice that the formal $1/m^{2}_{Q}$ suppression of the 
 spin-dependent interactions in Eq. (\ref{sdint}) is cancelled by the $m_{Q}$ dependence of the involved matrix elements. 
This result arises from our choice of the Bohr radius as the reference dimension scale, thus 
the average value $e.g.$ of $\langle 1/r^{3} \rangle_{nl}$ must be proportional to $1/a^{3}_{Q}\approx m^{3}_{Q} \alpha^{3}_{s}$. This is also
valid for the velocity-dependent terms in the spin-dependent sector not shown here.

Secondly, with this result, we get actually a consistent expansion in powers of $\alpha^{2}_{s}$ for the masses of heavy quarkonium. Al leading 
order we simply get the perturbative physics scale $\mu_{Q}=2m_{Q}$.  
At order $\alpha^{2}_{s}$, the coefficient is given by $\frac{4}{9}m_{Q}\epsilon_{nl}(\lambda)$ and involves non-perturbative QCD effects 
mimicked by the string tension $\lambda$. This coefficient is actually different for the different quarkonium configurations. The different coefficients 
are calculated exactly here and involve the analytic solution for the normalized energies of the Cornell potential, $\epsilon_{nl}(\lambda)$. 
For the quarkonium spectrum, the non-perturbative effects manifest at order $\alpha^{2}_{s}$ 
in the $\lambda$ dependence of $\epsilon_{nl}(\lambda)$, which is responsible for the breaking of the $l$-degeneracy of the $n$-levels of 
the Coulomb-like interaction.
The order $\alpha^{4}_{s}$ corrections include spin-independent and spin-dependent interactions and their non-perturbative  effects are 
calculated using Rayleigh-Schr\"{o}dinger perturbation theory. Our 
approximate calculation of these effects shows that the size of non perturbative effects at this order are modulated by the mean values 
of the space part of the corresponding operators which also involve the probabilities of the Cornell potential, $|R_{nl}(r,\lambda)|^{2}$. 

These results show that, for the heavy quarkonium spectrum, the suppression of higher order terms in the non-relativistic 
expansion (including non-perturbative effects) 
depends only on the expansion in $\alpha^{2}_{s}(m_{Q})$ and the whole expansion makes sense as long as $\alpha_{s}(m_{Q})$ is still small, 
which we expect to be satisfied for heavy quarkonium. Notice that Eq. (\ref{mnslj}) is consistent with the power counting rules of NRQCD, 
where we have a perturbative scale $m_{Q}$ and non-perturbative effects are suppressed 
by powers of $v^{2}\approx \alpha^{2}_{s}(m_{Q})$. In this concern, we remark that in Eq. (\ref{mnslj}) a new scale appear, given by
\begin{align}
\mu_{f}(m_{Q})&=\frac{64}{243} m_{Q} \alpha^{4}_{s},
\end{align}
which is the natural scale of the fine splittings of the heavy quarkonium spectrum. 

The mass hiearchies in the fine splittings of the measured heavy quarkonium spectrum noticed at the end of the previous section can be 
understood from Eq. (\ref{mnslj}) and the values of average spin-dependent operators in Table \ref{spinfac}. Indeed, notice that Eq. (\ref{mnslj}) 
can be written as
\begin{equation}
M[n^{2S+1}L_{J}] =M^{SI}_{nl}(\lambda) +M^{SD}[n^{2S+1}L_{J}](\lambda),
\end{equation}
where the spin-independent part, $M^{SI}_{nl}(\lambda)$, contain terms of order $\alpha^{2}_{s}$ and $\alpha^{4}_{s}$ which, for a given $n$, 
still preserve the $m$-degeneracy of the $l$-levels. This degeneracy is broken by the spin-dependent contribution 
$M^{SD}[n^{2S+1}L_{J}](\lambda)$ which only contains terms of order $\alpha^{4}_{s}$ producing the fine splittings of heavy quarkonium.

From the mean values of the spin operators in Table \ref{spinfac}, we can see that the spin-dependent contribution to the masses of the 
$l=0$ states in a given level $n$, 
get contributions of the spin-spin interactions only and we obtain 
\begin{equation}
M[n^{3}S_{1}] - M[n^{1}S_{0}]=   \mu_{f}~   f^{ss}_{nl}(\lambda) > 0,
\label{M3S1m1S0}
\end{equation}
thus the $n^{3}S_{1}$ states are heavier than the $n^{1}S_{0}$ states. This pattern is exhibited by the $n=1,2$ levels of heavy quarkonium as discussed 
at the end of the previous section and we predict this behavior to be valid for every level $n$ in both heavy quarkonium sectors.  

The ordering of the $^{3}P_{J}$ states in a given level $n$ can also be obtained from Eq. (\ref{mnslj}). Indeed, a straightforward calculation 
using the spin factors Table \ref{spinfac} yields
\begin{align}
M[n^{3}P_{2}] - M[n^{3}P_{1}]&=  \mu_{f}~  \left(   \frac{18}{5} f^{(-3)}_{n1} (\lambda) 
- \frac{3}{4} \lambda f^{(-1)}_{n1} (\lambda)   \right) \equiv \mu_{f} F^{P}_{21}(n,\lambda), \label{FPn21}\\
M[n^{3}P_{1}] - M[n^{3}P_{0}]&=  \mu_{f} ~ 
\left( \frac{9}{2} f^{(-3)}_{n1} (\lambda) - \frac{3}{8} \lambda f^{(-1)}_{n1} (\lambda)   \right) \equiv \mu_{f} F^{P}_{10}(n,\lambda). 
 \label{FPn10}
\end{align}

As for the $D$-waves in the $n$-th level, the mass splittings are given by 
\begin{align}
M[n^{3}D_{3}] - M[n^{1}D_{2}]&= \mu_{f} ~ \left(  \frac{81}{14} f^{(-3)}_{n2} (\lambda) 
-\frac{9}{8} \lambda f^{(-1)}_{n2} (\lambda)   \right)\equiv \mu_{f} F^{D}_{32}(n,\lambda),  \label{FDn32} \\
M[n^{3}D_{2}] - M[n^{3}D_{1}]&=  \mu_{f}  ~
 \left(  6 f^{(-3)}_{n2} (\lambda) - \frac{3}{4} \lambda f^{(-1)}_{n2} (\lambda)   \right)\equiv \mu_{f} F^{D}_{21}(n,\lambda). 
  \label{FDn21}
\end{align}
In Fig. (\ref{FPFD}) we plot the functions $F^{P}_{21}(n,\lambda), F^{P}_{10}(n,\lambda), F^{D}_{32}(n,\lambda),  F^{D}_{21}(n,\lambda)$ for $n=2,3$, 
calculated with the $[101/100]$ Pad\'{e} approximant, where we can see that these functions are positive. This results yields the ordering 
\begin{equation}
M[n^{3}P_{2}] >  M[n^{3}P_{1}] > M[n^{3}P_{0}],
\label{Porder}
\end{equation}
for $n=2,3$ states and
\begin{equation}
M[n^{3}D_{3}] >  M[n^{3}D_{2}] > M[n^{3}D_{1}] ,
\label{Dorder}
\end{equation}
for $n=3 $. This ordering is clearly seen in the physical spectrum of bottomonium for the 
$2^{3}P_{J}, 3^{3}P_{J}$ and $3^{3}D_{J}$ states. In the case of charmonium,
the measured $2^{3}P_{J}$ states also exhibit this ordering. For $n=3$, it requires the assignment of the $3^{3}D_{J}$ states to the 
$\psi(3842)$, $\psi(3823)$ and $\psi(3770)$ states. Notice that at the qualitative level discussed in this section, both the inverted 
spectrum and the order $\alpha^{4}_{s}$ corrections obtained here yields a consistent picture for the heavy quarkonium spectrum. 

\begin{figure}%
\centering
\includegraphics[width=7cm]{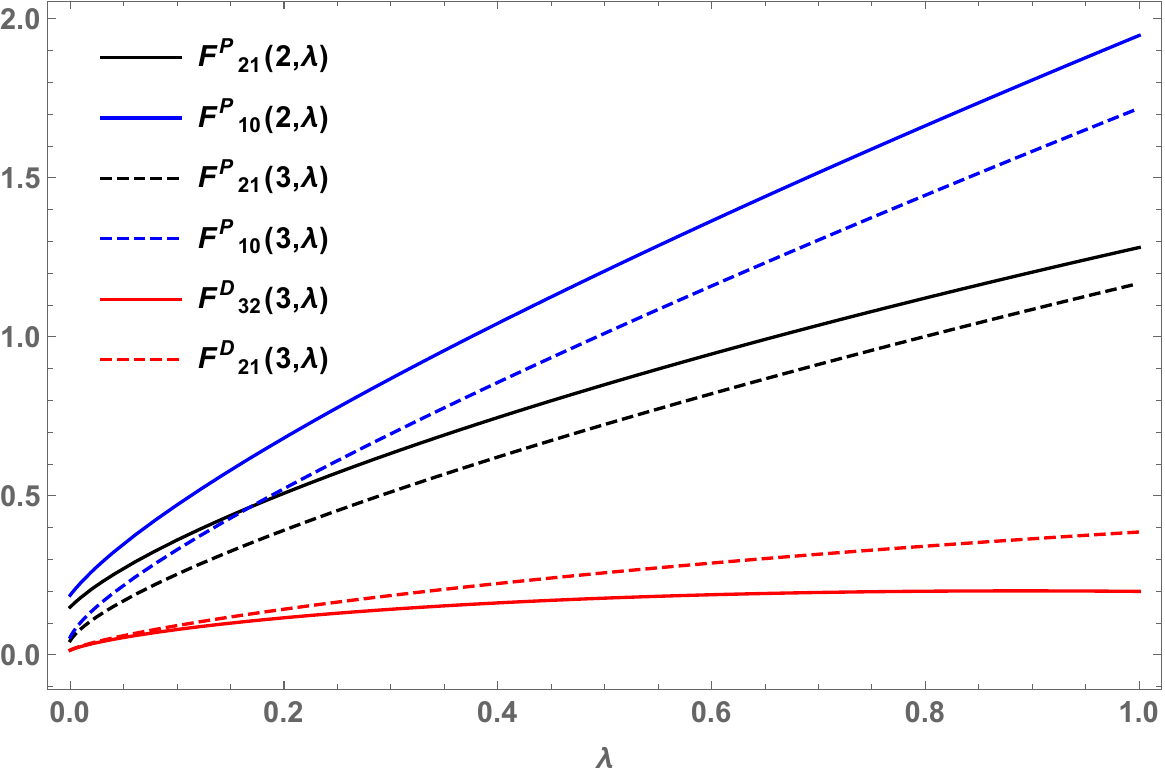} %
\caption{Combinations $F^{P}_{21}(2,\lambda), F^{P}_{10}(2,\lambda),F^{P}_{21}(3,\lambda), F^{P}_{10}(3,\lambda), 
F^{D}_{32}(2,\lambda),  F^{D}_{21}(3,\lambda)$ appearing in Eqs. (\ref{FPn21}, \ref{FPn10},\ref{FDn32},\ref{FDn21}) for $n=2,3$, as functions 
of $\lambda$.
}
 \label{FPFD}
\end{figure}

\section{Asessing non-perturbative effects and predictions for the missing heavy quarkonium states}
It is not the aim of the present work to give a detailed numerical analysis of the predictions of the complete analytical solution of the 
Cornell potential and the leading relativistic corrections obtained here for the quarkonium spectrum, but it would be interesting to have 
a more precise estimate of the physical normalized string tension $\lambda_{Q}$ than that given by the purely non-relativistic physics in 
Eq. (\ref{Bohrlambda}). In this section we will obtain it confidently 
from the fine splittings. It will allow us to make well defined predictions on the mass values of some missing states and 
on the range of mass values where other missing heavy quarkonium states must lie,
which is relevant for the experimental searches of these states. 

 In this concern, we can see from Eq. (\ref{M3S1m1S0}) that the fine splittings 
in the $S$-wave sectors are dictated by the fine energy scale $\mu_{f}$ and the value of $f^{ss}_{nl}(\lambda)$ at the physical value
$\lambda_{Q}$. Indeed, a straightforward calculation yields
\begin{align}
M[2^{3}S_{1}] - M[2^{1}S_{0}]&=  \mu_{f}(m_{Q}) f^{ss}_{20}(\lambda_{Q}), \\
M[1^{3}S_{1}] - M[1^{1}S_{0}]&= \mu_{f}(m_{Q}) f^{ss}_{10}(\lambda_{Q}).
\label{fsQ}
\end{align}
Notice that these splittings are not affected by spin-independent relativistic corrections which cancel in the difference. The fine scale $\mu_{f}(m_{Q})$
cancels in the ratio and a robust  extraction of the physical value of $\lambda_{Q}$ can be done solving
\begin{equation}
\frac{M[2^{3}S_{1}] - M[2^{1}S_{0}]}{M[1^{3}S_{1}] - M[1^{1}S_{0}]}=  \frac{f^{ss}_{20}(\lambda_{Q})}{f^{ss}_{10}(\lambda_{Q})}.
\label{lambdaQeq}
\end{equation}
It is important to remark that the normalized wave function at the origin, 
$f^{ss}_{n0}(m_{Q})$, actually receive considerable radiative corrections as noticed in \cite{Eichten:1980mw}. 
These corrections largely cancel in the ratio of the normalized wavefunctions in Eq. (\ref{lambdaQeq}) and a reliable estimate of $\lambda_{Q}$ for 
heavy quarkonium can be obtained from this equation, which is 
independent of the $\alpha_{s}$ and $m_{Q}$ factors. The ratio on the right hand side can be obtained from the complete 
analytical solution to the Cornell potential and depends only on $\lambda$. The experimental value of the left hand side of this equation 
can be extracted from Tables \ref{bbarbspec} and \ref{cbarcspec} and we get
\begin{align}
\frac{M[\Upsilon(2S)] - M[\eta_{b}(2S)] }{ M[\Upsilon(1S)] - M[\eta_{b}(1S)] }&= 0.395\pm 0.017 , \label{ratiob}\\
\frac{M[\psi(2S)] - M[\eta_{c}(2S)] }{ M[J/\psi(1S)] - M[\eta_{c}(1S)] }&= 0.429 \pm 0.008. \label{ratioc}
\end{align}

The dimensionless functions $f^{ss}_{nl}(\lambda)$ are related to $R^{2}_{nl}(0,\lambda)$ by Eq. (\ref{fssnl}). We reconstructed 
$R_{10}^{2}(0,\lambda)$ and $R_{20}^{2}(0,\lambda)$ using the $[151/150]$ and $[150/150]$ Pad\'{e} approximants. 
With these reconstructed functions, solving Eq. (\ref{lambdaQeq}) we get the following values for the 
physical normalized string tension at the bottomonium and charmonium mass scales
\begin{align}
\lambda_{b}=0.361\pm 0.054 , \qquad
\lambda_{c}=0.482 \pm 0.033 .
\label{l4}
\end{align}
Notice that these values are considerably lower than those extracted from the data on the average masses of the $n=1,l=0$ and $n=2,l=0,1$ 
in a purely non-relativistic model in Eq. (\ref{Bohrlambda}). The physical values in Eq. (\ref{l4}) incorporates the leading relativistic effects 
in a controlled $\alpha^{2}_{s}$ expansion. The uncertainties in the ${\cal{O}}(\alpha^{4}_{s})$ result in Eqs. (\ref{l4}) do not include theoretical 
uncertainties due to other ${\cal{O}}(\alpha^{6}_{s})$ terms or due to additional QCD effects, which must be eventually considered, but 
we expect these effects to be largely  cancelled in the ratios in Eqs. (\ref{ratiob},\ref{ratioc}).
 
Once fixed the physical values of $\lambda_{Q}$, from our complete analytical solution we can calculate the numerical factors 
$\epsilon_{nl}(\lambda_{Q})$ and  $f^{ss}_{nl}(\lambda_{Q})$ which are collected in Table \ref{values3332} for values up to $n=4$. 
To this end we use the $\epsilon_{nl}(\lambda)$ and $f^{ss}_{nl}(\lambda)$ functions reconstructed with the  $[151/150]$ and $[150/150]$  Pad\'{e} 
approximants. The actual value of the corresponding functions lie between these approximants and we estimate the
uncertainty in the reconstruction as the ratio of the difference and the sum of these functions. We list the reconstruction uncertainties
in the calculation of the physical quantities in the last column of Table \ref{values3332}.

\begin{table}[!ht]
    \begin{tabular}{|c|c|c|}
    \hline
   $Observable$ & $Value$ & $Rec. ~Uncertainty$\\
\hline
$\lambda_{b}$ & $ 0.361\pm 0.054$ & $  $\\
\hline
$\lambda_{c}$ & $ 0.482\pm 0.033$ & $ $\\
\hline
$\epsilon_{10}(\lambda_{b}) $ & $-0.518\pm 0.06 $ & $1.0 \times 10^{-16}$\\
\hline
$\epsilon_{20}(\lambda_{b}) $ & $1.124\pm 0.165 $ & $2.1\times 10^{-13}$ \\
\hline
$\epsilon_{30}(\lambda_{b}) $ & $2.070\pm 0.245 $ & $2.1 \times 10^{-7}$ \\
\hline
$\epsilon_{40}(\lambda_{b}) $ & $2.828\pm 0.314 $ & $6.8 \times 10^{-5} $\\
\hline
$\epsilon_{10}(\lambda_{c}) $ & $ -0.374\pm 0.039 $ & $7.4\times 10^{-17}$ \\
\hline
$\epsilon_{20}(\lambda_{c}) $ & $ 1.480\pm 0.094 $ & $1.1\times 10^{-11} $\\
\hline
$\epsilon_{30}(\lambda_{c}) $ & $ 2.596\pm 0.138 $ & $1.8\times 10^{-6}$ \\
\hline
$\epsilon_{40}(\lambda_{c}) $ & $ 3.500\pm 0.176 $ & $2.7\times 10^{-4}$ \\
\hline
$ f^{ss}_{10}(\lambda_{b})$ & $ 5.681\pm 0.212 $ & $7.8 \times 10^{-17} $ \\
\hline
$ f^{ss}_{20}(\lambda_{b})$ & $2.246\pm 0.182 $ & $2.9\times 10^{-13}$ \\
\hline
$ f^{ss}_{30}(\lambda_{b})$ & $ 1.681 \pm 0.155 $ & $2.8 \times 10^{-7}$ \\
\hline
$ f^{ss}_{40}(\lambda_{b})$ & $ 1.440\pm 0.141 $ & $8.0 \times 10^{-5}$ \\
\hline
$ f^{ss}_{10}(\lambda_{c})$ & $6.140\pm 0.121 $ &$<1\times 10^{-17}  $\\
\hline
$ f^{ss}_{20}(\lambda_{c})$ & $ 2.634\pm 0.101 $ & $ 1.5\times 10^{-11}$ \\
\hline
$ f^{ss}_{30}(\lambda_{c})$ & $ 2.014\pm 0.087 $& $ 2.2 \times 10^{-6} $\\
\hline
$ f^{ss}_{40}(\lambda_{c})$ & $1.743\pm 0.080  $ & $ 3.0 \times 10^{-4} $\\
\hline

\end{tabular}
\caption{Numerical values of observables relevant for the calculation of the spectrum of the $S$-wave states of heavy quarkonium, obtained from the 
eigenvalues and eigenstates of the Cornell potential, reconstructed with the $[151/150]$ Pad\'{e} approximant from the corresponding 
series calculated to order $\lambda^{301}$. }
\label{values3332}
\end{table}

\subsection{Bottomonium}
The fine structure scale for bottomonium can be obtained from Eq. (\ref{M3S1m1S0}}) which yields
\begin{equation}
\mu_{f}(m_{b})=\frac{M[\Upsilon(1S)]-M[\eta_{b}(1S)]}{f^{ss}_{10}(\lambda_{b})} =10.86 \pm 0.54~MeV.
\end{equation}

As a cross check, we obtain the same value if we use instead data on the $n=2$, $S$-wave states.

We can see in Table \ref{bbarbspec} that states in the $n=1,2$ levels are complete and follow the inverted spectrum 
pattern predicted by the Cornell potential and the $P$-waves ordering obtained from the leading relativistic corrections. 
For $n\geq3$  the states $\eta_{b}(nS)$ are missing. The value of the masses of these states
can be obtained from Eq. (\ref{M3S1m1S0}) as
\begin{equation}
 M[\eta_{b}(nS)]=M[\Upsilon(nS)] - \mu_{f}(m_{b})   f^{ss}_{n0}(\lambda_{b}).
\end{equation}
Using the values for $f^{ss}_{30}(\lambda_{b})$ and $f^{ss}_{40}(\lambda_{b})$ quoted in Table \ref{values3332}, and the experimental
values for the masses of the $\Upsilon(3S)$ and $\Upsilon(4S)$ in Table \ref{bbarbspec} we predict
the masses of the missing $\eta_{b}(3S)$ and $\eta_{b}(4S)$ as
\begin{align}
 M[\eta_{b}(3S)]&= 10337\pm 2 ~ MeV, \\
 M[\eta_{b}(4S)]&= 10564\pm 2 ~MeV.
\end{align}

As to the $D$-wave states in the $n=3$ level with configurations $3^{3}D_{3}$, $3 ^{1}D_{2}$, $3^{3}D_{2}$ and $3 ^{1}D_{1}$, 
only the $3^{3}D_{2}$ (the $\Upsilon_{2}(1D)$) has been discovered with a mass $M[\Upsilon_{2}(1D)]=10163.7 ~MeV$. The ordering 
of $D$-wave states in Eq. (\ref{Dorder}) and the inverted spectrum requires the masses of the $3^{3}D_{3}$, $3 ^{1}D_{2}$ 
 states to be in range $[10163, 10232]~MeV$. Similarly, the mass of the $3 ^{1}D_{1}$ 
state is predicted to be in the $[10023, 10163]~MeV$ range.

Concerning the states in the $n=4$ level of bottomonium, we have the following missing $P,D$ and $F$-wave states:  
$4^{1}P_{1}, 4^{3}P_{0},  4^{3}D_{3}, 4^{1}D_{2}, 4^{3}D_{2},4^{3}D_{1}, 4^{3}F_{4}, 4^{1}F_{3}, 4^{3}F_{3}, 4^{3}F_{2}$. 
Our complete analytical solution predicts all these states to be below the $\chi_{b2}(3P)$ 
state, and above the $\Upsilon(3S)$ state, i.e. in the energy range  $[10355,10524]MeV$, with $P$ states being heavier than the $D$ 
states which in turn are more massive than the $F$ states. In particular, the $F$ states must be around the mass of the $\Upsilon(3S)$ 
state because it is precisely with these levels where the crossing phenomena starts.

\subsection{Charmonium}
The values of the masses for charmonium states listed in Table \ref{cbarcspec} shows that states in the $n=1,2$ levels are complete 
and follow the inverted spectrum pattern predicted by the Cornell potential.  Also, the ordering of the $P$-waves follows the pattern 
predicted by the leading relativistic corrections. We can extract the value of the fine structure scale for charmonium from 
Eq. (\ref{M3S1m1S0}) and data on the $n=1$, $S$-wave states which yields
\begin{equation}
\mu_{f}(m_{c})=\frac{M[J/\psi(1S)-M[\eta_{c}(1S)]}{f^{ss}_{10}(\lambda_{c})}= 18.37 \pm 0.36~MeV.
\end{equation}
We cross checked this result using instead data of the $n=2$, $S$-wave charmonium states obtaining the same value.

In the $n=3$ level of charmonium the first challenge is the identification of the $3^{3}S_{1}$ charmonium state (the $\psi(3S)$). 
The calculation of the mass of 
this state requires the complete numerical analysis of Eq. (\ref{mnslj}) which is beyond the scope of this paper. However, 
modulo corrections of order $\alpha^{4}_{s}$, the Bohr scale can be obtained
from two successive Bohr levels, e.g. from the $2S$ and $1S$ bottomonium states we get
\begin{equation}
\mu_{B}(m_{b})\approx \frac{M[\Upsilon(2S)]-M[\Upsilon(1S)]}{\epsilon_{20}(\lambda_{b}) - \epsilon_{10}(\lambda_{b})}=343~MeV.
\end{equation}  
Similar values are obtained using the $3S$ and $2S$ levels
\begin{align}
\mu_{B}(m_{b})\approx \frac{M[\Upsilon(3S)]-M[\Upsilon(2S)]}{\epsilon_{30}(\lambda_{b}) - \epsilon_{20}(\lambda_{b})}=350~MeV.
\end{align}  
For the charmonium Bohr scale we get
\begin{equation}
\mu_{B}(m_{c})\approx \frac{M[\psi(2S)]-M[J/\psi(1S)]}{\epsilon_{20}(\lambda_{c}) - \epsilon_{10}(\lambda_{c})}=319~MeV.
\end{equation}  
We can use this value to estimate the mass of the $\psi(3S)$ state as
\begin{equation}
M[\psi(3S)]\approx M[\psi(2S)] + \mu_{B}(m_{c})(\epsilon_{30}(\lambda_{c}) - \epsilon_{20}(\lambda_{c}))= 4041 ~MeV.
\end{equation}
We expect order $\alpha^{4}_{s}$ corrections to these results, meaning corrections of the order of $\mu_{f}(m_{c})= 18.37 ~MeV$. 
Notice that the only $\psi$ state above the $\psi(2S)$ in Table \ref{cbarcspec} which is consistent with these values is the 
$\psi(4040)$.  We conclude that the $\psi(4040)$ is the $3^{3}S_{1}$ charmonium state. 
This conclusion is reinforced by the fact that the mass of the $3^{3}P_{2}$  (the $\chi_{c2}(3930)$) state is below the mass of the 
 $3^{3}S_{1}$ state and the $\psi(3770)$, $\psi(3823)$ and $\psi(3842)$ assumed as the $3^{3}D_{J}$ charmonium states lie 
 below the $S$ and $P$ waves as required by the inverted spectrum of the Cornell potential .

Once we identify the $\psi(4040)$ as the $\psi(3S)$ charmonium state, we are able to predict the mass of the $3^{1}S_{0}$ charmonium 
state (the $\eta_{c}(3S)$) which according to Eq. (\ref{M3S1m1S0}) must have a mass
\begin{equation}
M[\eta_{c}(3S)]= M[\psi(3S)] - \mu_{f}(m_{c})   f^{ss}_{30}(\lambda_{c})= 4003 \pm 4~ MeV.
\end{equation}

According to the inverted spectrum obtained with the complete analytical solution of the Cornell potential, the $n=3$, $P$-wave states must lie 
below the $\psi(3S)$ and the $\eta_{c}(3S)$. This pattern requires  the $\chi_{c2}(3930)$ to be the $\chi_{c2}(2P)$ state. The remaining 
$P$-waves must lie below the $\chi_{c2}(3930)$ and it is not likely that the $\chi_{c0}(3915)$ be a $\bar{c}c$ state because it is too close to the 
 $\chi_{c2}(3930)$. However, rigourous statements on the identification of the $n=3$ $P$-wave charmonium states with the measured 
$\chi_{c0}(3915)$, $\chi_{c1}(3872)$ and on the missing $h_{c}(2P)$ requires to do the complete numerical analysis of our main result in Eq. (\ref{mnslj})
including the complete set of spin-independent operators not considered here. 

\section{Conclusions and perspectives}

In this work we completely solve the Cornel potential using the supersymmetric expansion algorithm introduced in Ref. \cite{Napsuciale:2024yrf}.
The solutions are obtained in the form of power series in the normalized string tension $\lambda$. It is shown that the coefficients in the series 
satisfy algebraic recurrence relations which can be solved to the desired order. For these purposes, we wrote a Mathematica symbolic code, 
freely available upon request and calculate the series of the energy levels to order $\lambda^{301}$.
Although the power series have a small convergence radius, the actual value of the energy functions $E_{nl}(\lambda)$ and of the radial solutions 
$R_{nl}(r,\lambda)$ can be reconstructed from the power series using Pad\'{e} approximants. This reconstruction gets more faithful as we increase 
the number of terms calculated in the power series and calculations to order $\lambda^{301}$ allow us to reconstruct the levels up to $n=4$ with high 
precision. 

The first main result of the complete analytical solution obtained in this work is the prediction of an {\it{inverted 
spectrum}} for the Cornell potential. This means that energy levels depend on two quantum numbers, the principal quantum 
number $n$ and the orbital quantum number $l$, this dependence being actually a function of $n^{2}$ and $L^{2}\equiv l(l+1)$, and 
for a given $n$ the energy levels {\it{decrease}} with increasing $l$. The radial probabilities have the same shape as the Coulomb 
ones but the peaks are shifted to smaller radius, thus Cornell eigenstates are more compact than Coulomb eigenstates.

We calculate the critical values of $\lambda$ for each level, defined by  $\epsilon_{nl}(\lambda_{cr})=0$ for values of $n$ up to $n=9$. The energy levels 
exhibit the crossing phenomena starting with the $n=4$ level where the $\epsilon_{43}(\lambda)$ crosses with the $\epsilon_{30}(\lambda)$ 
level for $\lambda\approx 0.4$.

We apply this solution to the problem of the calculation of heavy quarkonium properties.  The heavy quarkonium masses depend on two 
well defined energy scales, the perturbative physics scale $\mu_{Q}=2m_{Q}$,
the natural scale for the Bohr-like levels $\mu_{B}=4m_{Q}\alpha^{2}_{s}/9$, and on the normalized energy solutions, $\epsilon_{nl}(\lambda)$.  
Collecting the masses of all the heavy quarkonium states considered as "well established" by the Particle Data Group we show that the so far 
measured heavy quarkonium masses exhibit the inverse spectrum pattern predicted by the Cornell potential.

A first estimate of the values of the parameters $\bar{\lambda}_{Q}$, $\bar{m}_{Q}$, $\bar{\alpha}_{s}(\bar{m}_{Q})$ is extracted considering 
that the Bohr-like levels correspond to the average values of the physical quarkonium masses for both $Q=b,c$ and using the experimental values 
for the lowest lying $\bar{M}_{10}$, $\bar{M}_{20}$, and $\bar{M}_{21}$. Using these values we predict the average masses for the highest levels. 
The only level with available data for the complete multiplet is the $n=3$,$l=1$ level. The prediction for the corresponding average mass, $\bar{M}_{31}$ 
agree with the predictions for the central value within $0.2\%$, which corresponds to $22 ~MeV$. 

Details of the structure of heavy quarkonium so far missing like the heavy quarkonium radius and the 
heavy quark squared velocities can be assessed from our solution. We find striking differences in these observables with respect to the Coulomb 
values. Indeed, a calculation of the inverse mean radius yields a result $\langle nl | (r/a_{Q})^{-1}|nl\rangle$ which 
differ for its Coulomb $1/n^{2}$ behaviour. The non-perturbative interactions mimicked by the linear term yield smaller radius than the Coulomb 
interaction for the $n$ levels, which in a given level $n$ increases with the value of the angular momentum $l$.
As to the average squared velocity we find $\langle nl | v^{2}|nl\rangle=C_{nl}(\lambda)\alpha^{2}_{s}$ 
as expected from $NRQCD$, but the proportionality constant is large compared with the Coulomb value $C_{nl}(0)=16/9n^{2}$ and the non-perturbative 
effects causes the squared velocity to depend on the orbital quantum number $l$ and to have a value around $2\alpha^{2}_{s}$ for all levels.
This result signals to the relevance of relativistic corrections for all the levels of heavy quarkonium, not only for the ground state. 

We consider then the leading relativistic corrections and calculate the corresponding fine structure splittings using Rayleigh-Schr\"{o}dinger 
perturbation theory. The second main result of this calculation is the expression of the masses of the $n^{2S+1}L_{J}$ heavy quarkonium 
states as an expansion in $\alpha^{2}_{s}(m_{Q})$. Indeed, the formal $m^{-2}_{Q}$ suppression of the relativistic corrections is actually cancelled 
by $m^{2}_{Q}$ factor from the involved matrix elements (which include the non-perturbative effects of the linear term) leaving actually a power series 
in $\alpha^{2}_{s}$. The spectrum involves now three well defined energy scales: The perturbative scale given by $\mu_{Q}(m_{Q})=2m_{Q}$, 
the scale dictating the Bohr-like levels, $\mu_{B}=\frac{4}{9}m_{Q}\alpha^{2}_{s}(m_{Q})$, and the scale of the fine splittings 
$\mu_{f}(m_{Q})= \frac{64}{243}m_{Q}\alpha^{4}_{s}(m_{Q})$. The 
masses of heavy quarkonium states are given in terms these scales, the normalized energy solutions $\epsilon_{nl}(\lambda)$ and the average values 
of powers of the heavy quarkonium radius normalized to the Bohr radius $a_{Q}$ which are dimensionless functions $f^{(k)}(\lambda)$ depending 
only on the normalized string tension $\lambda$.

The first general prediction arising from this calculation is that the $n^{3}S_{1}$ states are heavier than the $n^{1}S_{0}$ states. Also, we conclude that 
the masses of the $^{3}P_{J}$-wave states have a well defined ordering given by 
$M[n^{3}P_{2}]> M[n^{3}P_{1}] > M[n^{3}P_{0}]$ for $n=2,3$.  This ordering is clearly exhibited by the measured states in the 
$n=2,3$ levels of bottomonium and by the measured states in $n=2$ level of charmonium. Similarly, the masses of the $^{3}D_{J}$ states of heavy 
quarkonium are predicted to satisfy the following hierarchy:  $M[n^{3}D_{3}] > M[n^{3}D_{2}] > M[n^{3}D_{1}]$ for $n=2,3$. 

A confident extraction of the physical value of the normalized string tension $\lambda_{Q}$ for heavy quarkonium is done from data 
on the fine splittings of the lowest lying $l=0$ states, finding $\lambda_{b}= 0.361\pm 0.054$ and $\lambda_{c}=0.482\pm 0.033$. These values 
and the obtained solutions allow us to fix the fine splitting heavy quarkonium scales to $\mu_{f}(m_{b})= 10.86 \pm 0.54 ~MeV$ and 
$\mu_{f}(m_{c})= 18.37 \pm 0.36 ~MeV$. With these values and the obtained normalized probabilities evaluated at the origin, we are 
able to predict the values of the masses of the $\eta_{b}(3S)$ and $\eta_{b}(4S)$ states.
 
An estimate of the Bohr scales from data of the $n=1, 2$ $S$-wave states yields $\mu_{B}(m_{b})\approx 343~MeV$ and 
$\mu_{B}(m_{c})\approx 319~MeV$. These results 
allow us to identify the $\psi(4040)$ state as the ${3}^{3}S_{1}$
(the $\psi(3S)$) charmonium state and to conclude that the measured masses of the $\psi(3842)$, $\psi(3823)$ and  $\psi(3770)$ are 
consistent with the identification of these states as the $3^{3}D_{3}$, $3^{3}D_{2}$ and $3^{3}D_{1}$ $\bar{c}c$ states respectively. 

In the 
$n=3$ $P$-wave sector we identify the $\chi_{c2}(3930)$ as the $3^{3}P_{2}$ charmonium state ($\chi_{c2}(2P)$) and conclude that the 
remaining $P$-wave states ($h_{c}(2P), \chi_{c1}(2P), \chi_{c0}(2P)$ must lie in the $[3842,3925]~MeV$ energy range with 
the $P$-wave states being heavier that the $D$-wave states and the $^{3}P_{J}$ and $^{3}D_{J}$ ordered according to:
$M[3^{3}P_{2}] >  M[3^{3}P_{1}] > M[3^{3}P_{0}]>M[3^{3}D_{3}] >  M[3^{3}D_{2}] > M[3^{3}D_{1}] $.

Finally, this paper yields another interesting application of the supersymmetric expansion algorithm to a long standing unsolved potential of primary 
importance in the phenomenological description of non-perturbative effects in heavy quarkonium physics. The complete solution of this problem allow us 
 to go from the perturbative to the non-perturbative regime in a controlled manner, revealing and quantifying the role of non-perturbative effects 
 in the conformation of heavy quarkonium. There remain 
many possibilities for the use of our complete analytical solution to the Cornell potential, including the estimate of fundamental quantities 
arising in the $NRQCD$ and $pNRQCD$ effective theories and the complete phenomenological analysis of the leading relativistic corrections. 

\section{Acknowledgments}
We thank Prof. Franz F. Schoberl for providing us with the Mathematica code of Ref. \cite{Lucha:1998xc}. We used it at an early stage
of this work to confirm numerically the inverted spectrum predicted by the Cornell potential. 
A.E. Villanueva-Guti\'{e}rrez acknowledges financial support from CONACyT M\'{e}xico under a Level III  Researcher Assistantship.

\bibliography{SEA_Cornell}
\bibliographystyle{JHEP}

\end{document}